\documentclass[11pt]{article}

\usepackage{geometry}
\usepackage{amsmath}
\usepackage{amsthm}
\usepackage{siunitx}
\usepackage[small]{titlesec}
\usepackage{enumitem}
\usepackage{subcaption}
\usepackage{booktabs}
\usepackage[hidelinks]{hyperref}
\usepackage{amsfonts}
\usepackage{tikz}
\theoremstyle{definition}
\newtheorem{corollary}{Corollary}
\newtheorem{theorem}{Theorem}
\newtheorem{proposition}{Proposition}
\newtheorem{problem}{Problem}
\newtheorem*{remark}{Remark}
\newtheorem{definition}{Definition}
\newtheorem{lemma}{Lemma}
\usepackage[numbers,super,sort&compress]{natbib}
\setcitestyle{square,numbers}
\renewcommand\citenum[1]{\cite{#1}}
\let\OLDthebibliography\thebibliography
\renewcommand\thebibliography[1]{
  \OLDthebibliography{#1}
  \setlength{\parskip}{0pt}
  \setlength{\itemsep}{0pt plus 0.3ex}
}
\makeatletter
\renewcommand\@biblabel[1]{#1.}
\makeatother
\def\abstract{\normalfont%
      \textit{\bf Abstract}---\,}%
 
\usepackage{moreverb}

\newcommand{\Ri}[1]{\mathbb{R}^{#1}}
\newcommand{\Rip}{\mathbb{R}_+}
\newcommand{\Comp}[1]{\mathbb{C}^{#1}}
\newcommand{\cont}[1]{\mathcal{C}^{#1}}
\newcommand{\I}[1]{{{I}}_{#1}}
\newcommand{\0}[1]{{{ 0}}_{#1}}

\DeclareMathOperator*{\argmin}{arg\,min}
\newcommand{\sgn}{\text{sgn}}
\newcommand{\sat}{\text{sat}}
\newcommand{\distance}{\text{dist}}
\newcommand{\inv}{^{-1}}
\newcommand{\pinv}{^{\dagger}}
\newcommand{\transp}{^{\mathsf{T}}}
\newcommand{\rank}[1]{{\text{rank}~{ #1}}}
\newcommand{\vspan}[1]{{\text{span}~{ #1}}}

\newcommand{\lipschitz}{{l}}
\newcommand{\bigO}{\mathcal{O}}
\newcommand{\jac}[1]{D{#1}}

\newcommand{\nom}{\star}
\newcommand{\nomorb}{{\eta}_\nom}
\newcommand{\orbneighb}{\mathcal{N}}
\newcommand{\sysstate}{{{x}}}
\newcommand{\varstate}{\delta\sysstate}
\newcommand{\linstate}{{{y}}}
\newcommand{\gc}{{{q}}} 
\newcommand{\gcs}{{{q}}}
\newcommand{\Dgc}{{\dot{\gc}}}

\newcommand{\DDgc}{{\ddot{\gc}}}

\newcommand{\ac}{{{u}}}

\newcommand{\nfb}{k}
\newcommand{\lfb}{K}
\newcommand{\dist}{\Delta}
\newcommand{\smf}{\sigma}
\newcommand{\SM}{\Sigma}
\newcommand{\vc}{{{\Phi}}}
\newcommand{\vcs}{{{\phi}}}
\newcommand{\mgvc}{\theta}
\newcommand{\mg}{s}
\newcommand{\mgspace}{\mathcal{I}}
\newcommand{\nvel}{\rho}
\newcommand{\nomflow}{\mathcal{F}}
\newcommand{\tvc}{{{x}_{\perp}}}
\newcommand{\tvcjac}{D\tvc}
\newcommand{\mono}{{\mathcal{M}}}
\newcommand{\stm}{{\Psi}}
\newcommand{\xs}{{{x}_{\mg}}}
\newcommand{\prj}{p}
\newcommand{\prjspace}{\mathcal{X}}
\newcommand{\Dp}{{D}\prj}

\begin{document}
%%%%%%%%%%%%%%%%%%%%
%%%% Title page %%%%
\author{ Christian Fredrik Sætre$^{1,\dag}$  $\ |$ Anton S. Shiriaev$^{1,2}$ $\ |$   Leonid B. Freidovich$^{3,2}$ \\[0.1cm]  Sergei V. Gusev$^2$ $\ |$ Leonid M. Fridman$^{4,2}$}

\title{\large \bf Robust Orbital Stabilization: A Floquet Theory--based Approach\thanks{This research was supported by the
Research Council of Norway, grant number 262363; by 
CONACyT (Consejo Nacional de Ciencia y Tecnologia),  project 282013; and by PAPIIT-UNAM (Programa de Apoyo a Proyectos de Investigacion e
Innovacion Tecnologica) under Grant IN 106622.
\newline
{\indent $^\dag$Correspondence:} Christian Fredrik Sætre, O. S. Bragstads plass 2D, Elektroblokk D, 1. etg., Gløshaugen, Trondheim, Norway.  Email: {christian.f.satre@\{ntnu.no, gmail.com\}} }}

\date{\small $^1$Department of Engineering Cybernetics, NTNU, Trondheim, Norway.
\\
$^2$Department of Information Technologies and AI,
Sirius Univ. of Science and Technology, Sochi,
Russia.
\\
$^3$Department of Applied Physics and Electronics, Umeå University, Umeå, Sweden.
\\
$^4$Departamento de Ingeniería de Control y Robótica,  UNAM, Mexico City, Mexico.
}

\maketitle

%%%%%%%%%%%%%%%%%%
%%%% Abstract %%%%
\begin{abstract}{The design of robust orbitally stabilizing feedback is considered. From a known orbitally stabilizing controller for a nominal, disturbance-free system, a robustifying  feedback extension is designed utilizing the sliding-mode control (SMC) methodology. The main contribution of the paper is to provide a constructive procedure for designing the time-invariant switching function used in the SMC synthesis. More specifically, its zero-level set (the sliding manifold) is designed using a real Floquet--Lyapunov transformation to locally correspond to an invariant subspace  of the Monodromy matrix of a transverse linearization. This ensures  asymptotic stability of the periodic orbit when the system is confined to the sliding manifold, despite any system uncertainties and external disturbances satisfying a matching condition. The challenging task of oscillation control of the underactuated Cart-Pendulum system subject to both matched- and unmatched disturbances/uncertainties demonstrates the efficacy of the proposed scheme.
}
\\[0.2cm] 
\end{abstract}

\maketitle

%%%% Introduction %%%%
\section{Introduction}
\emph{Orbital stabilization} is the utilization of {time-invariant} feedback to generate an asymptotically stable limit cycle in the {resulting autonomous} closed-loop system, corresponding to the closed orbit of a desired periodic motion \cite{fradkov1998introduction}. It is a natural way of phrasing the task of feedback design in applications where  the desired operating mode is oscillatory, and it can have several benefits compared to trajectory tracking methods,  especially for nonlinear systems which are underactuated and non-minimum phase. For instance, it makes stability analysis a far more tractable problem as the {closed-loop system is time-invariant}, as opposed to time-varying. Moreover, it lessens the burden on the control action as it does not need to ensure the ``timing'' of the motion,  in the sense that the system's states do not need to converge to the desired orbit in phase with a time-varying reference trajectory.

{There exist several methods for designing orbitally stabilizing feedback for different classes of systems in the literature; see, for example, references  \citenum{banaszuk1995feedback,hauser1995maneuver,mohammadi2018dynamic,ortega2018orbital,saetre2019excessive,saetre2020excessive,shiriaev2005constructive,shiriaev2010transverse,yi2020orbital}, to name but a few. These methods share a primary goal: to simultaneously generate and stabilize  self-induced oscillations via continuous feedback. This means that a precise mathematical model of the system to be controlled is generally required for these methods to be successfully applied. Indeed, since the resulting closed-loop system is autonomous, any unknown disturbance or model discrepancies (e.g., due to unmodelled dynamics or uncertain parameters), may significantly alter its behaviour. Thus, if not taken into consideration, unknown perturbations can  result in a change of both the shape and location of the induced limit cycle, or even lead to the instability of the desired motion.
Yet, with the exception of a few methods that are either only applicable for a very limiting class of systems \cite{park2009orbital} or only ensure asymptotic orbital stability of some of the system's states \cite{santiesteban2008second}, most orbital stabilization methods are not designed  specifically with robustness in mind.  }

This lack of robustness can be problematic, as uncertainty and unknown disturbances will often be an inherent part of many such tasks. For instance, in  dynamic manipulation of compliant objects, say, rolling an orange on the palm of a robotic hand, trying to accurately model all the complex phenomena of the contact between the interacting objects is not only a daunting task, it will often be infeasible in practice.
A more realistic strategy is to instead use a model which is ``good enough'' to generate an approximate motion and to design a nominal feedback for it, whereas the remaining disturbances and uncertainties are lumped together and compensated for by a robustifying feedback extension.\footnote{Although not part of the focus of this paper, it may often be advantageous to complement the robust feedback with an adaptive scheme to estimate uncertain parameters.}

 In regards to the design of such an extension, the sliding mode control (SMC)  methodology, with its well-known insensitivity to bounded {perturbations} satisfying a matching condition \cite{shtessel2014sliding}, is especially well suited.  It consists of two  main steps: 1) the construction of a \emph{switching function}, whose zero-level set defines a \emph{sliding manifold} on which the system has desired properties; and 2) the design of a control law which ensures that the sliding manifold is reached in finite time despite of any matched {perturbations}. 

There exist a large array of different strategies \cite{gutman1979uncertain,utkin2013sliding,nagesh2014multivariable} to solve the latter problem provided that the switching surface  is given. Thus the question we are looking to answer in regards to robust orbital stabilization is  the following: How to construct a time-invariant switching function that defines a sliding manifold upon which the system's states  converges to a desired orbit? The main contribution of this paper is to provide a new constructive procedure for designing such a function. This procedure is applicable for a large class of nonlinear systems, including underactuated mechanical systems. 

We begin by stating a detailed problem formulation and provide an outline  of the proposed solution in the next section.

%%%% Notation %%%%
\paragraph{Notation.}
 $\Ri{}$ denotes the reals and $\Rip{}$ the nonnegative reals. $\I{n}$ denotes the $n\times n$ identity matrix, while $\0{n\times m}$ is a $n\times m$ matrix of zeros, with $\0{n}=\0{n\times n}$. 
Given two column vectors $w$ and $v$,  the shorthand notation $[w;v]=[w\transp,v\transp]\transp$ is used. For any $z= [z_1;z_2\dots;z_n]\in\Ri{n}$,   $\|z\|_p:=\left( \sum_{i=1}^n|z_i|^p\right)^{\frac{1}{p}}$ denotes the $p$-norm on $\Ri{n}$, with $\|z\|=\|z\|_2$   the  Euclidean norm. For a smooth mapping $h:\Ri{n}\to\Ri{m}$, we denote by $\jac{h}(\sysstate)=[\frac{\partial h}{\partial x_1}(\sysstate),\dots,\frac{\partial h}{\partial x_n}(\sysstate)]$ its Jacobian matrix at $\sysstate\in\Ri{n}$.   $\|h(\sysstate)\|=\bigO(\|\sysstate\|^{k})$ if $\|h({\sysstate})\|\le c\|{\sysstate}\|^k$ as $\|\sysstate\|\to 0$ for some $c>0$.  $A\pinv\in\Ri{m\times n}$ denotes the pseudo- (Moore--Penrose) inverse of a  full rank matrix $A\in\Ri{n\times m}$. Positive semi-definite (PSD) and -definite (PD) matrices are denoted by $\succeq 0$ and $\succ 0$, respectively. The eigenvalues of $A\in\Ri{n\times n}$ with the smallest and largest real part are denoted by $\lambda_{\min}(A)$ and $\lambda_{\max}(A)$, respectively. Given $M\in\Comp{n\times n}$, we denote by  $\overline{M}$ its (element-wise) complex conjugate. For $a\in\Ri{}$,  $\sgn(a)\in\{-1,0,1\}$  and  $\sat(a)\in[-1,1]$ denote the signum- and saturation function, respectively.

%%%%%%%%%%%%%%%%%%%%%%%%%%%
%%%% Problem statement %%%%
\section{Problem Formulation and Outline of the Proposed Solution}\label{sec:ProbForm}
\subsection{Problem Formulation}
Consider a nonlinear control-affine system with an  {unknown, matched perturbation:}
\begin{equation}\label{eq:NonLinSys}
    \dot{\sysstate}=f(\sysstate)+g(\sysstate)\left(\ac+\dist(\sysstate,t)\right). 
\end{equation}
Here $\sysstate(t)\in\Ri{n}$  denotes the state at time $t\in\Rip$, $\ac(t) \in\Ri{m}$ represents the control inputs, $m<n$, while the perturbation term  $\dist:\Ri{n}\times \Rip\to\Ri{m}$, consisting of  system uncertainties and unknown external disturbances,  has a  known upper bound:\footnote{While the restrictions upon $\dist(\cdot)$ are here taken to be quite conservative for simplicity, they can be somewhat relaxed. For example, the proposed scheme can be easily extended to  a disturbance term of the form $\dist:\Ri{n}\times \Ri{m}\times\Rip\to\Ri{m}$ for which $\|\dist(\sysstate,\ac,t)\|\le \dist_0+\dist_u\|u\|+\alpha(\sysstate,t)$  is satisfied given known constants $\dist_0,\dist_u\in\Rip$ and a known function $\alpha:\Ri{n}\times \Rip\to \Rip$.}  $\|\dist\|\le \dist_M$ for all $\sysstate\in\Ri{n}$ and $t\in\Rip$.
We assume  $f:\Ri{n}\to\Ri{n}$ to be  $\mathcal{C}^2$ (twice continuously differentiable) and that the columns of $g(\cdot)\in\Ri{n\times m}$, denoted $g_i:\Ri{n}\to\Ri{n}$, $i\in\{1,\dots,m\}$,  are linearly independent and (locally) Lipschitz continuous.

{It will be assumed that a  bounded, $T$-periodic solution $\sysstate_{\nom}(t)=\sysstate_{\nom}(t+T)$
  of the nominal (i.e. perturbation-free) and undriven (i.e. $\ac\equiv 0$) system is known for some $T>0$;} that is $\dot{\sysstate}_\nom(t)=f(\sysstate_\nom(t))$ and $0< \|f(\sysstate_\nom(t))\|<\infty$
for all $t\in\Rip$. {It will further be assumed} that a $\mathcal{C}^2$-mapping $\nfb:\Ri{n}\to\Ri{m}$ is known, satisfying $\nfb(\sysstate_\nom(t))\equiv 0$,  which (locally) renders  $\sysstate_\nom(t)$  an {\emph{exponentially orbitally (Poincaré) stable}}\cite{fradkov1998introduction,khalil2002nonlinear} solution of the nominal  closed-loop system,  described by
\begin{equation}\label{eq:NomNonLinSys}
    \dot{\chi}=f(\chi)+g(\chi)\nfb(\chi), \quad \chi\in\Ri{n}.
\end{equation}
That is to say,  if
\begin{equation}\label{eq:NomOrb}
    \nomorb:=\big\{\sysstate\in\Ri{n}| \ \sysstate=\sysstate_\nom(t), \ t\in[0,T)\big\}
\end{equation}
denotes the corresponding periodic \emph{orbit},
and if we define the distance  $\distance(\sysstate,\nomorb):=\inf_{y\in\nomorb}\|x-y\|$, then this stability concept is understood in the following sense.
\begin{definition}[Orbital stability]
{
A solution $\sysstate_\nom(\cdot)$ (resp. its orbit $\nomorb$) of the autonomous system  \eqref{eq:NomNonLinSys}  is said to be \emph{orbitally  stable} (resp. stable) if, for every $\epsilon>0$, there is a $\delta=\delta(\epsilon)>0$, such that for any solution $\sysstate(\cdot)$ of \eqref{eq:NomNonLinSys} satisfying $\distance(\sysstate(t_0),\nomorb)<\delta$, it is implied that $\distance(\sysstate(t),\nomorb)<\epsilon$ for all $t\ge t_0$. The solution (resp. its orbit) is said to be \emph{asymptotically orbitally stable} (resp. asymptotically stable), if it is orbitally stable (resp. stable), and there is an open tubular neighbourhood $\orbneighb(r):=\big\{\sysstate\in\Ri{n}: \ \distance(\sysstate,\nomorb)<r\big\}$ of the orbit $\nomorb$ for some $r>0$, such that $\sysstate(t_0)\in\orbneighb(r)$ implies $  \distance(\sysstate(t),\nomorb)\to 0$ as $t\to\infty$. The solution  (resp. its orbit) is said to be \emph{exponentially orbitally stable} (resp. exponentially stable) if there exist   constants $\lambda,C>0$, such that $\distance(\sysstate(t),\nomorb)\le C e^{-\lambda (t-t_0)}$ for all $t\ge t_0$.
}
\end{definition}

The fact that $\nfb(\cdot)$ renders $\sysstate_\nom(\cdot)$ an exponentially orbitally  stable solution of the disturbance-free system \eqref{eq:NomNonLinSys} does of course  in no way guarantee that it will also be an (asymptotically) orbitally stable solution of \eqref{eq:NonLinSys} in the presence of the matched perturbation given by $g(\sysstate)\dist(\sysstate,t)$. {In fact, it may no longer be a solution of the closed-loop system at all.} For this reason, we consider the task of utilizing the knowledge of $\nfb(\cdot)$ to instead design a \emph{robust} controller which also renders $\sysstate_\nom(\cdot)$ an   asymptotically orbitally stable  solution   of the  system \eqref{eq:NonLinSys}. We do this by searching for a time-invariant \emph{switching function}, whose zero-level set defines a \emph{sliding manifold/surface}  {upon} which all solutions sufficiently close to the desired orbit converges to it. More precisely, we are looking to solve the following problem using our prior knowledge of the {nominal} feedback $\nfb(\cdot)$.
%%%% Main Problem %%%%
\begin{problem}\label{prob:MainProblem}
Find a time-invariant, $\mathcal{C}^2$  \emph{switching function} $\smf:\Ri{n}\to\Ri{m}$ such that if restricted to the \emph{sliding manifold}, defined by
\begin{equation}\label{eq:SM}
    \SM:=\{\sysstate\in\Ri{n}: \ \smf(\sysstate)=\0{m\times 1}\},
\end{equation}
then, for any $\sysstate$ within some tubular neighbourhood of the orbit $\nomorb$,  the system \eqref{eq:NonLinSys} experiences the \emph{equivalent control}  \cite{utkin2013sliding} 
\begin{equation}\label{eq:EqCont}
    \ac_{eq}=\hat{\nfb}(\sysstate)-\dist(\sysstate,t)
\end{equation}
where $\hat{\nfb}:\Ri{n}\to\Ri{m}$ is a $\mathcal{C}^1$ mapping   {satisfying $\hat{\nfb}(y)\equiv 0$ and $\jac{\hat{\nfb}}(y)=\jac{{\nfb}}(y)$ for all $y\in\nomorb$. That is,  the first-order approximations of $\hat{\nfb}(\cdot)$ and  $\nfb(\cdot)$ along $\nomorb$ are equal. }
\end{problem}

\begin{remark}
    The use of Utkin's  \emph{equivalent control method} \cite{utkin2013sliding}  ensures that the control ``experienced'' by the system when confined to the manifold \eqref{eq:SM} corresponds to \eqref{eq:EqCont}, which is both disturbance rejecting and asymptotically orbitally stabilizing. Thus any  motion (of reduced order) of the system \eqref{eq:NonLinSys} in sliding mode may be considered to evolve as if $\dot{\sysstate}=f(\sysstate)+g(\sysstate)\hat{\nfb}(\sysstate)$. The mapping $\hat{\nfb}(\cdot)$ is considered rather than the known feedback $\nfb(\cdot)$  as it allows for an added level of flexibility in the design of the switching function (adding or removing higher order terms), but still keeping the local orbitally stabilizing feedback properties.
\end{remark}

Notice also the absence of an explicit form of a sliding mode control law in Problem~\ref{prob:MainProblem}. Indeed, as previously stated, the main focus of this paper is not the design of sliding mode controllers  per se, but rather the design of sliding manifolds on which the orbit \eqref{eq:NomOrb} is asymptotically stable.
Of course, if such a  manifold is given, then some sort of sliding mode control law (be that a relay-type, unit-vector, higher-order, etc.) is necessary in order to bring the system's states onto it in finite time. While the choice of such a control law is important in regards to aspects such as, for example, chattering attenuation and the required assumptions upon the unknown disturbance term, it  does not affect the corresponding equivalent control \eqref{eq:EqCont}, which instead is completely determined  by the choice of switching function. Hence the tasks of designing and stabilizing the corresponding sliding manifold may be considered separately, with our focus in this paper mainly on the former. 

%%%%%%%%%%%%%%%%%
%%%% Outline %%%%
\subsection{Outline of the Proposed Solution}\label{sec:outline}
 It is well known that the local (general) behavior of a smooth nonlinear system about one of its hyperbolic equilibrium points can be determined by the corresponding (Jacobian) linearization about this point (the Hartman--Grobman theorem). In particular, the invariant subspaces of the linearization correspond to locally invariant manifolds of the nonlinear system (cf. Theorem 6.1 in \citenum{hartman2002ordinary}).   This local equivalence may also be utilized for the purpose of designing robust controllers for nonlinear systems subject to matched disturbances. For instance, one can attempt to design the switching function used in the SMC synthesis such that its zero-level set locally corresponds to a stable invariant subspaces of the linearization of a nominal model of the system about the desired hyperbolic equilibrium.\footnote{Some further complementary comments regarding the use of invariant subspaces of a linearization for the design of switching functions are provided in Appendix~\ref{app:InvariantManifolds},}  
 
 The  approach we suggest in this paper for solving Problem~\ref{prob:MainProblem} is based on similar ideas. Namely  on the local equivalence between the  stable  invariant manifolds of the nominal (nonlinear) closed-loop system \eqref{eq:NomNonLinSys}
and the  stable  {invariant} subspaces of the corresponding first-order approximation system along the orbit \eqref{eq:NomOrb}. 
How to utilize these ideas as to construct a time-invariant switching function $\smf(\cdot)$ may  not be immediately obvious, however. Indeed,  linearizing  \eqref{eq:NomNonLinSys} along the solution $\sysstate_\nom(t)$ just results in a time-varying (periodic) system which evidently will have the non-vanishing solution $\dot{\sysstate}_\nom(t)$. This implies that one must find a real invariant subspace of appropriate dimension among the remaining $(n-1)$ independent solutions. However, any annihilator of such a subspace will be time-varying in general.  
Thus, as to obtain a solution to Problem~\ref{prob:MainProblem},  one first needs to  construct such a subspace, and then, more importantly,  design from its annihilator the time-invariant switching function $\smf(\cdot)$. 
The main contribution of this paper is to provide a constructive procedure for doing so. 

More specifically, we suggest for this purpose the following three-step approach:\footnote{This idea is inspired by the method proposed by Freidovich and Gusev \cite{freid2020} in regards to a specific procedure for mechanical systems. The  approach in this paper builds upon and generalizes their ideas, as well as expand their applicability to a larger class of systems  by providing  a constructive procedure for obtaining solutions to Problem~\ref{prob:MainProblem}.  } 

\begin{itemize}
    \item[\bf 1)] \emph{Transverse linearization:} Derive the linear periodic system  corresponding to the first approximation (linearization) along the nominal orbit of the dynamics of a set of $(n-1)$ \emph{transverse coordinates}, whose origin correspond to the nominal orbit;
    
    \item[\bf 2)] \emph{Floquet--Lyapunov transformation:} Transform this linear periodic  system into a linear time-invariant  system through a real Floquet--Lyapunov factorization of its state transition matrix;
    
    \item[\bf 3)] \emph{Invariant subspace-based switching function design:} Construct a switching function for this linear time-invariant system, corresponding to an annihilator of one of its  real invariant subspaces whose co-dimension equals the number of controls. 
\end{itemize}

\paragraph{Outline}
This three-step approach is presented in a top-to-bottom way through the next three sections:  We first demonstrate how to use invariant subspaces to design switching functions for stabilizing the origin of linear time-invariant systems in Sec.~\ref{sec.LTI}, corresponding to step 3) above. Then in Sec.~\ref{sec:LTP} the same is done for  linear time-periodic systems using Floquet--Lyapunov transformations, which is used in step 2). We then solve Problem~\ref{prob:MainProblem}  directly in Sec.~\ref{sec:NonLinSysTVC} for nonlinear systems of the form \eqref{eq:NonLinSys} using a set of transverse coordinates and  the linearization of their dynamics, with the paper's main result stated in Section~\ref{sec:MR}.

The remainder of the paper is then organized as follows. A suggestion for a simple unit vector-based sliding mode control law for the nonlinear system is given in Sec.~ \ref{sec:ContDes}, which is briefly compared to a Lyapunov redesign based controller in Sec.~\ref{sec:LRC}. Then the concrete task of stabilizing upright oscillations of the cart-pendulum system subject to both matched- and unmatched uncertainties is considered as an illustrative example in Section~\ref{sec:CPmainSec}. Lastly, we state some concluding remarks and possible direction for further work in Sec.~\ref{sec:conclusion}. 

Note that some supplementary material  which may be useful for implementing the proposed procedure is provided in Appendix~\ref{app:SupMat}. Also note that the proofs of all the statements  in this paper, except those in Section~\ref{sec.LTI}, are found in Appendix~\ref{app:proofs}.  

%%%%%%%%%%%%%%%%%%%%%
%%%% LTI systems %%%%
\section{Invariant Subspace-based Switching Function Design for LTI Systems}\label{sec.LTI}
In  this section, we will show how  invariant subspaces can be used to construct switching functions for linear time-invariant (LTI) systems {with matching perturbations:}
\begin{equation}\label{eq:LTI}
    \dot{\linstate}=A\linstate+B\left(\ac+\dist(\linstate,t)\right), \quad \linstate \in\Ri{\bar{n}}, \quad \ac\in\Ri{\bar{m}}.
\end{equation}
Here $A\in\Ri{\bar{n}\times \bar{n}}$ is constant,  $B\in\Ri{\bar{n}\times \bar{m}}$  is of full rank and $\dist(\cdot)\in\Ri{\bar{m}}$ is unknown, but has a known upper bound as before. 

Under the assumption of the stabilizability of the pair $(A,B)$,  we will now consider the following task.
%%%% LTI Problem %%%%
\begin{problem}\label{prbolem:GenSSforLTI}
Given a matrix $\lfb\in\Ri{\bar{m}\times \bar{n}}$ such that $A^{cl}:=A+B\lfb$ is a Hurwitz (stable) matrix, find a full rank matrix $S\in\Ri{\bar{m}\times \bar{n}}$ such that when restricted to the manifold
\begin{equation}\label{eq:SlidingManifold}
    \SM:=\{\linstate\in\Ri{\bar{n}}:\  \smf(\linstate):=S\linstate\equiv 0\}
\end{equation}
the system \eqref{eq:LTI} experiences the equivalent control $\ac_{eq}=\lfb\linstate-\dist(\linstate,t)$.
\end{problem}

It is important to note that Problem~\ref{prbolem:GenSSforLTI} of course differs from Problem~\ref{prob:MainProblem} in the sense that it considers the stabilization of the origin of \eqref{eq:LTI}, whereas Problem~\ref{prob:MainProblem}  considers the task of stabilizing a (non-trivial) periodic orbit. The motivation for this problem is nevertheless the same, namely the design of a sliding manifold on which the stability of the nominal closed-loop system is preserved and  matched perturbations are rejected. 

\begin{lemma}\label{lemma:SM_LTI}
    If $S\in\Ri{\bar{m} \times \bar{n}}$ is such that  
    \begin{enumerate}
        \item $\det(SB) \neq 0$,
        \item  $S\linstate =0$ $\implies SA^{cl}\linstate =0$,
    \end{enumerate}
    then it is a solution to Problem~\ref{prbolem:GenSSforLTI}. 
\end{lemma}
\begin{proof}
    {Following the equivalent control approach,\cite{utkin2013sliding} we set $S\dot{\linstate}^* \equiv 0$ when $\linstate^*$ is confined to the sliding manifold \eqref{eq:SlidingManifold} to obtain $S\left[A\linstate^*+B\left(\ac_{eq}+\dist \right) \right]\equiv 0$.} By adding and subtracting $B\lfb\linstate^*$ inside the brackets, this can be equivalently rewritten as
    \begin{equation*}
         S\left[A^{cl}\linstate^*+B\left(\ac_{eq}+\dist-\lfb\linstate^* \right) \right]\equiv 0.
    \end{equation*}
    Since  $S\linstate^*=0$ implies $SA^{cl}\linstate^* =0$ (condition 2.) and the square matrix  $SB$ is nonsingular (condition 1.),   the above equality must correspond to the unique equivalent control $\ac_{eq}=\lfb\linstate^*-\dist$. Hence \eqref{eq:LTI} evolves as if $\dot{\linstate}^*=A^{cl}\linstate^*$
    when in sliding mode. 
\end{proof}

The fact that such a solution $S$ must be nonsingular (condition 1.) and be such that  $SA^{cl}y=0$ if $Sy=0$ (condition 2.) implies that  $S$ must be a left annihilator of a real  invariant subspace of $A^{cl}$ (see, e.g., Appendix~\ref{app:GenIS} for more details).
The  existence of such a matrix $S$ therefore boils down to the existence of such a subspace, for  which there are three obvious possibilities: 
\begin{enumerate}
    \item[S1.] There does not exist any real invariant subspace of $A^{cl}$ satisfying the conditions of the Lemma, that is, either  no subspace of codimension $\bar{m}$ or  $\rank{SB}<\bar{m}$ for any annihilator; 
    
    \item[S2.] There exists exactly one subspace of codimension $\bar{m}$  satisfying the conditions of the Lemma; 
    
    \item[S3.] There exist more than one such subspace.
\end{enumerate}

It is important to note that there is no guarantee that such a subspace will exist in general for an arbitrary stabilizing matrix $\lfb$. Thus, in the case of situation S1, one is forced to either find an alternative feedback matrix $K$,  use alternative methods to construct  $S$ directly\footnote{Knowledge of a stabilizing matrix $\lfb$ is of course not needed for constructing a sliding manifold for LTI systems. Indeed,  there exist several well-known approaches for designing the matrix $S$ directly; see e.g. Chapter 2.2 in  \citenum{shtessel2014sliding} or Chapter 7 in  \citenum{utkin2013sliding}.},    design a robustifying feedback extension utilizing other approaches (e.g., through Lyapunov redesign techniques \cite{khalil2002nonlinear,corless1981continuous,gutman1979uncertain}) or to use dynamic methods such as integral sliding mode control \cite{shtessel2014sliding}. 

Having the possibility to  choose a particular surface among many, as in situation S3, is of course the  most desirable.  Indeed, this  provides one with the possibility to pick a  subspace having certain properties, such as choosing the subspace which has the fastest convergence (that whose largest (negative) exponent has the largest magnitude). This also provides motivation for utilizing this approach beyond just for robustification purposes, in the sense that it can also be used to drive the system onto a prespecified subspace having some desired properties.

%%%%%%%%%%%%%%%%%%%%
%%%% SMC design %%%%
\subsection{Sliding mode control design for reaching surfaces constructed based on Lemma~\ref{lemma:SM_LTI}}
Should a sliding surface satisfying Lemma~\ref{lemma:SM_LTI} be found,  then the next step is to design some feedback controller which ensures that the sliding manifold \eqref{eq:SlidingManifold} is reached in finite time. For both the sake of completeness and to  motivate the control design we propose for the nonlinear system in Section ~\ref{sec:ContDes}, we provide {the following statements.}
%%%%%%%%%%%%%%%%%%%%%%%%%
\begin{lemma}\label{lemma:linCLsys}
    Let $S\in\Ri{\bar{m}\times \bar{n}}$ satisfy Lemma~\ref{lemma:SM_LTI} and suppose $\ac$ is taken as
    \begin{equation}\label{eq:LinCLsys}
        \ac=\lfb\linstate+(SB)\inv v
\end{equation}
in \eqref{eq:LTI}  for some $ v\in\Ri{\bar{m}}$.
Then the dynamics of $\sigma:=S\linstate$ {outside of} the sliding manifold $\SM$ are given by
\begin{equation}\label{eq:sigmaLinDyn}
    \dot{\sigma}=\mathcal{A}_\smf \smf+v+SB\dist(\linstate,t),
\end{equation}
where the constant matrix $\mathcal{A}_\smf:=SA^{cl}S\pinv$ is Hurwitz.
\end{lemma}
\begin{proof}
    Firstly, we may always write $\linstate=S\pinv\sigma+(\I{\bar{n}}-S\pinv S)\linstate$. Here  $S\pinv$ is taken as the unique Moore--Penrose pseudoinverse of $S$, i.e. $SS\pinv=\I{\bar{m}}$, although  any full-rank right-inverse may be used instead. Using this in $\dot{\smf}=S\dot{y}$ together with the fact that $SA^{cl}(\I{\bar{n}}-S\pinv S)\linstate\equiv \0{\bar{m}\times 1}$ for all $\linstate\in\Ri{n}$ if $S$ satisfies Lemma~\ref{lemma:SM_LTI}, one obtains \eqref{eq:sigmaLinDyn}  by inserting \eqref{eq:LinCLsys} into \eqref{eq:LTI}.
    
    Secondly, since $S$ annihilates a stable invariant subspace of $A^{cl}$, spanned by a set of its (real) generalized eigenvectors, the matrix  $\mathcal{A}_\smf\in\Ri{\bar{m}\times \bar{m}}$ is necessarily Hurwitz, with its spectrum  a subset of the spectrum of $A^{cl}$.
     Indeed, if $S_\perp\in\Ri{\bar{n}\times (\bar{n}-\bar{m})}$ is a basis of $\ker\{S\}$, then there exists a nonsingular matrix $Y\in\Ri{\bar{m}\times \bar{m}}$, a possibly singular matrix $Z\in\Ri{(\bar{n}-\bar{m})\times \bar{m}}$ and a  block diagonal Hurwitz matrix $\Lambda\in\Ri{\bar{n}\times \bar{n}}$ such that $A^{cl}=V\Lambda V\inv$ is a real Jordan form \cite[Ch. 3.4]{horn2012matrix} of $A^{cl}$, with $V=[S_\perp,S_\perp Z+S\pinv Y]$ and $V\inv=[(S_\perp\pinv-ZY\inv S)\transp;(Y\inv S)\transp]$. By partitioning $\Lambda$ as
    \begin{equation*}
        \Lambda=\begin{bmatrix}\Lambda_{11} & \Lambda_{12} \\
        \Lambda_{21} & \Lambda_{22} \end{bmatrix}, \quad \Lambda_{11}\in\Ri{(\bar{n}-\bar{m})\times(\bar{n}-\bar{m})}, \Lambda_{12}\in\Ri{(\bar{n}-\bar{m})\times \bar{m}}, \ \Lambda_{21}\in\Ri{\bar{m} \times(\bar{n}-\bar{m})}, \Lambda_{22}\in\Ri{\bar{m}\times \bar{m}},
    \end{equation*}
    one can show that $\mathcal{A}_\sigma=SA^{cl}S\pinv=Y[\Lambda_{22}-\Lambda_{21}Z]Y\inv$. Due to the specific structure of the real Jordan form and the fact that $S_\perp$ spans an invariant subspace of $A^{cl}$, namely $S_\perp=S_\perp\Lambda_{11}$, we must here have $\Lambda_{21}\equiv\0{\bar{m}\times (\bar{n}-\bar{m})}$. Hence the eigenvalues of $\mathcal{A}_\sigma$ are the eigenvalues of $\Lambda_{22}$, which in turn correspond to a subset of the spectrum of $A^{cl}$.
\end{proof}

%%%%%%%%%%%%%%%%%%%%%%
There exist several control strategies in the literature which may here be used to ensure that the origin of \eqref{eq:sigmaLinDyn} is reached in finite time despite of the perturbation $\dist$. 
As an example of such a controller, we provide the following  unit-vector approach \cite{shtessel2014sliding}. 

%%%%%%%%%%%%%%%%%%%%%%%
\begin{proposition}\label{prop:LinCL}
    Let $\mathcal{A}_\smf:=SA^{cl}S\pinv$ be as  in Lemma~\ref{lemma:linCLsys} and let $P=P\transp\in\Ri{\bar{m}\times \bar{m}}$ be the unique positive definite (PD) solution  to the Lyapunov equation $\mathcal{A}_\sigma\transp P+P\mathcal{A}_\sigma=-Q$ for some symmetric PD matrix $Q\in\Ri{\bar{m}\times \bar{m}}$ . Then  the control law \eqref{eq:LinCLsys} with
    \begin{equation}\label{eq:LinCL}
        v= {\begin{cases}-\mu \frac{\sigma}{\|\sigma\|} \quad &\text{if} \quad \|\sigma\| \neq0,  \\
        \quad 0 \quad &\text{if} \quad \|\sigma\| = 0 ,
        \end{cases}}  
        \qquad \text{for some} \quad    \mu \ge \frac{1}{\lambda_{min}(P)}\left[\frac{1}{2}\mu_\star+\lambda_{max}(P)\|SB\|\dist_M\right], \quad \mu_\star>0,
    \end{equation}
    guarantees that the sliding manifold \eqref{eq:SlidingManifold} is reached in finite time.
\end{proposition}
\begin{proof}
    It is well known that  $\mathcal{A}_\sigma$ being Hurwitz guarantees the existence of a unique solution $\Ri{\bar{m}\times \bar{m}}\ni P=P\transp\succ0$  to the Lyapunov equation \cite{khalil2002nonlinear}.
    Consider, therefore, the Lyapunov function candidate $V_\sigma:=\sigma\transp P \sigma$, such that by \eqref{eq:sigmaLinDyn},
    \begin{align*}
        \frac{d}{dt}V_\sigma&=\sigma\transp\left(\mathcal{A}_\sigma\transp P+P\mathcal{A}_\sigma-\frac{2\mu}{\|\sigma\|}P\right)\sigma+2\sigma\transp PSB\dist(\linstate,t) \\
        &\le -\lambda_{min}(Q)\|\sigma\|^2+2\left[\lambda_{max}(P)\|SB\|\dist_M-\mu\lambda_{min}(P)\right]\|\sigma\|.
    \end{align*}
    From the lower bound of $\mu$ one consequently obtains  $\frac{d}{dt}V_\sigma \le -\alpha V_\sigma -\beta \sqrt{V_\sigma}$ with $\alpha:=\frac{\lambda_{min}(Q)}{\lambda_{max}(P)}$ and $\beta:=\mu_\star/\sqrt{\lambda_{max}(P)}$. Using standard argument (see, e.g.,  Ch. 14.1.1 in
    \citenum{khalil2002nonlinear}) it can therefore be concluded that the sliding manifold $\SM$ is reached in finite time, {with the settling time $t_s$ satisfying the inequality  $t_s\le{2}{\alpha\inv}\ln\left({\alpha\beta\inv\sqrt{V_\sigma(0)}+1}{}\right) $ ; see Reference \citenum{yu2005continuous}.}
\end{proof}

We remark that, while the well-known chattering effect \cite{utkin2013sliding,shtessel2014sliding} is the main drawback of the controller \eqref{eq:LinCL}, methods for alleviating and attenuating this effect  using continuous approximations of \eqref{eq:LinCL} do exist; see, for example,  \citenum{khalil2002nonlinear,gutman1979uncertain,corless1981continuous}.  Although note that these methods only ensure convergence to a boundary layer of the sliding manifold. Alternatively,   the structure of \eqref{lemma:linCLsys}  may also allow (depending on the disturbance)  for the  possibility of utilizing multivariable super-twisting algorithms \cite{nagesh2014multivariable,lopez2019generalised}.

%%%%%%%%%%%%%%%%%%%%%
%%%% LTP systems %%%%
\section{Linear Periodic Systems and  Floquet--Lyapunov Transformations}\label{sec:LTP}
Consider now  a linear time-periodic (LTP) system
\begin{equation}\label{eq:LTP}
    \dot{\linstate}=A(t)\linstate+B(t)\left(\ac+\dist(\linstate,t)\right), \ t\in\Rip, \  \linstate\in\Ri{\bar{n}}, \ \ac\in\Ri{\bar{m}},
\end{equation}
with continuous, bounded, $T$-periodic matrix functions $A(t)=A(t+T)$ and $B(t)=B(t+T)$ of  minimal period $T>0$. As before,  $\dist(\cdot)\in\Ri{\bar{m}}$ is  unknown but has known upper bound $\dist_M$.

In a similar manner to the LTI systems in the previous section, let us assume that  a continuous, $T$-periodic matrix function $\lfb:\Rip\to\Ri{\bar{m}\times \bar{n}}$ is known such that the origin of the disturbance-free closed-loop system, given by
\begin{equation}\label{eq:LTP_CL}
    \dot{\chi}=A^{cl}(t)\chi, \quad A^{cl}(t):=A(t)+B(t)\lfb(t), 
\end{equation}
is exponentially stable. Letting $\stm_{A^{cl}}(\cdot)\in\Ri{\bar{n}\times \bar{n}}$ denote the \emph{state-transition matrix} (STM), i.e. the unique solution to
\begin{equation}\label{eq:STM}
    \frac{d}{dt}{\stm}_{A^{cl}}(t,t_0)=A^{cl}(t)\stm_{A^{cl}}(t,t_0), \ \ \stm_{A^{cl}}(t_0,t_0)=\I{\bar{n}},
\end{equation} 
then it is well known that this is equivalent to all the eigenvalues of the \emph{Monodromy matrix} 
\begin{equation}\label{eq:Monodromy}
\mono_{A^{cl}}:={\stm}_{A^{cl}}(T,0)    
\end{equation}
having magnitudes strictly less than one. 

Assuming knowledge of such a matrix $\lfb(\cdot)$, we will  in this section consider the following  Problem.
\begin{problem}\label{prbolem:GenSSforLTP}
    Find a $\mathcal{C}^1$ matrix function $S:\Rip\to\Ri{\bar{m}\times \bar{n}}$,  such that the forward invariance of the relation $S(t) \linstate(t)\equiv 0$ for all $t\ge t_0$ corresponds to the system \eqref{eq:LTP} experiencing the equivalent control $\ac_{eq}(t)=\lfb(t)\linstate(t)-\dist(\linstate(t),t)$ for all $t\ge t_0$.
\end{problem}

Although  this problem is naturally  more challenging than Problem~\ref{prbolem:GenSSforLTI} as the matrix $S(\cdot)$ might be time-varying (periodic), solutions can be found using the knowledge of the state-transition matrix.
%%%%%%% General solution %%%%%
\begin{lemma}\label{lemma:LTPstatement}
    Let $X_0\in\Ri{\bar{n}\times (\bar{n}-\bar{m})}$ be of full rank and suppose the  $\cont{1}$ matrix function $S:\Rip\to\Ri{\bar{m}\times \bar{n}}$   is a left annihilator of the range space of ${\stm}_{A^{cl}}(t,0)X_0$ at time $t$, that is $\|S(t){\stm}_{A^{cl}}(t,0)X_0p\|\equiv 0$ for all $p\in\Ri{(\bar{n}-\bar{m})}$ and any $t\ge0$. Then $S(t)$ is a solution to Problem~\ref{prbolem:GenSSforLTP} if $\rank[S(t)B(t)]=\bar{m}$ for all  $t\in\Rip$. Moreover, if  $S(t)$ is to be $T$-periodic, i.e. $S(t)=S(t+T)$ for any $t\ge 0$, then $X_0$ must be a basis of an invariant subspace of $\mono_{A^{cl}}$ of codimension $\bar{m}$.
\end{lemma}

\noindent 
\textit{Proof.} See Appendix~\ref{proof:lemma:LTPstatement}.

The question of how to generate and numerically construct such a matrix function $S(\cdot)$ therefore arises. For this purpose, suppose we can smoothly transform the  LTP system \eqref{eq:LTP} into an LTI one. This would allow us to readily use the theory outlined in the previous section, in particular Lemma~\ref{lemma:SM_LTI}. We demonstrate how this can be achieved  utilizing a \emph{Floquet--Lyapunov} (FL) transformation next.

%%%%%%%%%%%%%%%%%%%%%%%%%%%%%%%%%%%%%
\subsection{Floquet--Lyapunov Transformations }\label{sec:FLtrans}
Let $A:\Rip{}\to\Ri{\bar{n}\times \bar{n}}$ be a bounded and  continuous matrix function and consider the linear time-varying (LTV) system:
\begin{equation}\label{eq:LTPol}
    \dot{\linstate}=A(t)\linstate , \quad \linstate \in \Ri{n}, \quad t\in\Rip{}.
\end{equation}
Denote by $\stm_A(\cdot)$  the STM, i.e. $\linstate(t)=\stm_A(t,\tau)\linstate(\tau)$ for all $t,\tau\in\Rip$ (see \eqref{eq:STM}), and suppose there exists a real, constant, $\bar{n}\times \bar{n}$ matrix $F$ and a nonsingular, $\mathcal{C}^1$ matrix function $L:\Rip{}\to\Ri{\bar{n}\times \bar{n}}$ such that $\stm_A(\cdot)$ can be factorized as follows:
\begin{equation}\label{eq:FLfact}
    \Psi_A(t,0)=L(t)e^{Ft} \quad \forall t\in\Rip.
\end{equation}
The induced coordinate transformation $\linstate(t)=L(t)z(t)$ is then said to be a (real) \emph{Lyapunov transformation}, while the system \eqref{eq:LTPol} is said to be \emph{real-reducible}, in the sense that  $\dot z=Fz$ is time-invariant.

While it is well known that not all LTV systems are reducible,  Floquet \cite{floquet1883equations} demonstrated  that all linear time-periodic (LTP) systems are.
Thus, if  $A(t)$ is $T$-periodic and $L(t)=L(t+cT)$ for some integer $c$, then  $\linstate(t)=L(t)z(t)$ is referred to as a  $cT$-periodic  \emph{Floquet--Lyapunov} (FL) \emph{transformation}, while \eqref{eq:FLfact} will be referred to as a $cT$-periodic FL \emph{ factorization}. 

In the following, we will therefore take the matrix $A(\cdot)\in\Ri{\bar{n}\times \bar{n}}$  in \eqref{eq:LTPol} to be both continuous and  $T$-periodic, that is $A(t)=A(t+T)$. It is  known  that a real, $2T$-periodic FL  factorization always exists for LTP systems of the form \eqref{eq:LTPol} \cite{yakubovich1975,montagnier2003real,zhou2008classification}. The following statement demonstrates this fact (see also Theorem  3.1 in  \citenum{montagnier2003real} for a generalization of this theorem.)
\begin{theorem}[Real FL Transformation \cite{yakubovich1975}] \label{th:realFLtrans}
     If $A(t)=A(t+T)$ in \eqref{eq:LTPol} is continuous, then there always exists a real, continuously differentiable, nonsingular matrix function $L(t)$, as well as real, commuting matrices $F$ and $Y$, i.e. $FY=YF$, satisfying \begin{equation*}
         L(t+2T)=L(t),\qquad  L(t+T)=L(t)Y, \qquad Y^2=\I{\bar{n}},
     \end{equation*}
    such that \eqref{eq:FLfact} holds for the LTP  system \eqref{eq:LTPol}.
\end{theorem} 

The existence of real, $T$-periodic FL  factorizations,  however, depends upon the spectrum of the Monodromy matrix $\mono_A$. That is to say, since any such   factorization \eqref{eq:FLfact} naturally must satisfy
\begin{equation}\label{eq:MonodromyAndFLfact}
    \mono_A=\stm_A(T,0)=e^{FT},
\end{equation}
the existence of a real matrix $F$ is dependent on the existence of a real (matrix) logarithm of $\mono_A$, i.e. $\log \mono_A=FT$. Using this, together with the fact that $\stm_A(t,0)$ is nonsingular for all $t\in\Rip$, the following statement is just a well-known, straightforward consequence of Theorem 1 in  \citenum{culver1966existence}.

\begin{lemma}\label{lemma:existenceOfRealLog}
    The LTP system \eqref{eq:LTPol} has a real, $T$-periodic FL factorization of the form \eqref{eq:FLfact}  if, and only if, each Jordan block corresponding to an eigenvalue of $\mono_A$ with negative real part  appears an even number of times. 
\end{lemma}

Note that some methods for obtaining  FL factorizations are briefly discussed in Appendix~\ref{app:consFLfac}.

%%%%%%%%%%%%%%%%%%%%%%%%%%%%%%%%%%%%%%
\subsection{Switching surface design for LTP systems}\label{sec:LTPssd}
The next statement demonstrates how an FL factorization can be used to solve Problem~\ref{prbolem:GenSSforLTP}.
\begin{proposition}\label{prop:SMdesignLTP}
Let the pair $(L(t),F)$ be a real, $cT$-periodic FL factorization of the closed-loop system \eqref{eq:LTP_CL} for some positive integer $c$, and suppose the matrix $F$ has a real invariant subspace $\Lambda$ of co-dimension $\bar{m}$, that is $F\Lambda\subseteq \Lambda$. Then the matrix function $S(t):=\hat{S}L\inv(t)$ is a solution to Problem~\ref{prbolem:GenSSforLTP} if $\rank [S(t)B(t)]=\bar{m}$ for all $t\in[0,cT)$ and  $\hat{S}\in\Ri{
\bar{m}\times \bar{n}}$ is a full rank left-annihilator of $\Lambda$, that is $\hat{S}z=0$ for all $z\in\Lambda$.
\end{proposition}
\noindent 
\textit{Proof.} See Appendix~\ref{proof:prop:SMdesignLTP}.

\begin{remark}
This statement may easily be extended to any real-reducible linear time-varying system.
\end{remark}
Recall from Theorem~\ref{th:realFLtrans} that a real FL  factorization always exists for $c=2$, whereas the existence of a $T$-periodic factorization follows from Lemma~\ref{lemma:existenceOfRealLog}. The following statements demonstrates that a $2T$-periodic factorization may  result in a $T$-periodic $S(t)$.
\begin{corollary}
    \label{corr:SyS}
    Let the triplet $(L(t),F,Y)$ denote a real, $2T$-periodic FL factorization of the closed-loop system \eqref{eq:LTP_CL} as in Theorem~\ref{th:realFLtrans}  and let the conditions of Proposition~\ref{prop:SMdesignLTP} hold. Then the matrix function $S(t):=\hat{S}L\inv(t)$ is  $T$-periodic if $\hat{S}=\hat{S}Y$.
\end{corollary}
This is just a consequence of the fact that $L\inv(T)=Y$. For $\hat{S}=\hat{S}Y$ to be satisfied, however, it is clear that the rows of $\hat{S}$ must be linear combinations of the left eigenvectors of $Y$ corresponding to its unitary eigenvalues. This may of course also sometimes be possible even  when $Y\neq I_{\bar{n}}$ as   $Y^2=\I{\bar{n}}$ (the matrix $Y$ is involutory), and hence all its eigenvalues satisfy $\lambda_Y^2=1$. 

%%%%%%%%%%%%%%%%%%%%%%%%%%
%%%% Nonlinear system %%%%
\section{Sliding Manifold Design for Nonlinear Systems}\label{sec:NonLinSysTVC}
By taking inspiration from the statements in the previous sections, we now turn our attention back to the nonlinear system \eqref{eq:NonLinSys} and to Problem~\ref{prob:MainProblem}. In this regard, we begin by defining the following continuous and bounded, $T$-periodic matrix functions:
\begin{equation*}
    A(t):=\jac{f}(\sysstate_\nom(t)),    \quad   B(t):=g(\sysstate_\nom(t)), \quad \lfb(t):=\jac{k}(\sysstate_\nom(t)) \quad  \text{and}  \quad A^{cl}(t):=A(t)+B(t)\lfb(t).
\end{equation*}
Moreover, we let the corresponding state-transition matrix (STM) $\stm_{A^{cl}}(\cdot)$ and Monodromy matrix $\mono_{A^{cl}}$ be defined according to \eqref{eq:STM} and \eqref{eq:Monodromy}. 
{Let us also briefly recall the purpose behind Problem~\ref{prob:MainProblem}: To find a sliding variable $\smf:\Ri{n}\to\Ri{m}$ such that when in sliding mode sufficiently close to $\nomorb$, the equivalent control is given by $\ac_{eq}=\hat{\nfb}(\sysstate)-\dist(\sysstate,t)$, where $D\hat{\nfb}(\sysstate_\nom(t))\equiv \lfb(t) $ for all $t\in[0,T)$. To see why this is desirable, we insert the equivalent control into the dynamical system \eqref{eq:NonLinSys} to obtain
\begin{equation*}
    \dot{\sysstate}_\smf=f(\sysstate_\smf)+g(\sysstate_\smf)\hat{\nfb}(\sysstate_\smf).
\end{equation*}
This is the so-called \emph{ideal sliding equation}, whose first-order approximation system along the solution $\sysstate_\nom(\cdot)$ is necessarily equivalent to that of the  nominal system \eqref{eq:NomNonLinSys}, namely $\frac{d}{dt}\varstate=A^{cl}(t) \varstate$.
 By the Andronov--Vitt theorem\cite{fradkov1998introduction,leonov2006generalization}, it therefore follows that $\sysstate_\nom(\cdot)$ is an asymptotically orbitally stable solution of the ideal sliding mode equation.  As we can tie the solutions of this system to those of \eqref{eq:NonLinSys} when in sliding mode using Utkin's equivalent control method\cite{utkin2013sliding},  the following statement can be concluded.
 \begin{lemma}
    Let $\smf:\Ri{n}\to\Ri{m}$ be a solution to Problem~\ref{prob:MainProblem}. Then there is exists a tubular neighbourhood $\orbneighb$ of the nominal orbit  $\nomorb$,  such that if the states of the system \eqref{eq:NonLinSys} are restricted to the sliding manifold $\SM:=\left\{\sysstate\in\Ri{n}: \ \smf(\sysstate)=\0{m\times 1}\right\}$ within $\orbneighb$, then all solutions of \eqref{eq:NonLinSys} converges to $\nomorb$, or equivalently,  $\sysstate_\nom(\cdot)$ is rendered asymptotically orbitally stable therein.
 \end{lemma}
 }

The following conditions upon such a solution to Problem~1 may then be stated.
\begin{lemma}\label{lemma:IntroStatement}
    Let  $\sigma:\Ri{n}\to\Ri{m}$ be $\mathcal{C}^2$  and define the $T$-periodic matrix function $S(t):=\jac{\sigma}(\sysstate_\nom(t))$.     If $\sigma(\sysstate_\nom(t))\equiv 0,$  as well as 
    \begin{enumerate}
        \item  $\det[S(t)B(t)]\neq 0 $,
        \item $S(t)x=0$ $\implies$ $\left[\dot{S}(t)+S(t)A^{cl}(t)\right]\sysstate\equiv 0,$
    \end{enumerate}
      are satisfied for all $t\in[0,T)$, then $\sigma(\cdot)$ is a solution to Problem~\ref{prob:MainProblem}.
\end{lemma}
\noindent 
\textit{Proof.} See Appendix~\ref{proof:lemma:IntroStatement}.
%%%%%%%%%%%%%%%%%%%%%%%%%%%%%%%%

%%%%%%%%%%%%%%%%%%%%%%%%%%%%%%%%
\begin{remark}
It is not difficult to see that  condition 2. together with the fact that $S(t) =\jac{\smf}(\sysstate_\nom(t)) $ must be $T$-periodic, implies by Lemma~\ref{lemma:LTPstatement} that $S(t)$ must  be a left-annihilator of the range space of $\stm_{A^{cl}}(t,0) X_0$, with $X_0$ a basis of a real invariant subspace of the Monodromy matrix $\mono_{A^{cl}}$ of codimension $m$. This has an important implication: only the eigenvalues of the Monodromy matrix that correspond to its subspace with basis $X_0$ need to have magnitude less than one. Hence there may exist a sliding manifold on which all solution are asymptotically orbitally stable even though the feedback $\nfb(\cdot)$ is not fully orbitally stabilizing but asymptotically stabilizes a particular subspace of the first-order approximation system.
\end{remark}

Although Lemma~\ref{lemma:IntroStatement} provides sufficient conditions   for a mapping to be a solution to Problem~\ref{prob:MainProblem}, it does not provide a constructive procedure for obtaining it. We therefore demonstrate next how one can design such a switching function using a Floquet--Lyapunov transformation, by first  transforming the orbital stabilization problem into the task of stabilizing an origin through a change to so-called \emph{transverse coordinates}.

%%%%%%%%%%%%%%%%%%%%%%%%%%
\subsection{Preliminaries: Transverse coordinates and projection operators }\label{sec:MRprelims}
{Let the curve  $\xs:\mgspace\to\nomorb$, with $\mgspace:=[0,\mg_T)$, denote  a $\mathcal{C}^2$-smooth regular parameterization of  the orbit \eqref{eq:NomOrb}, that is}
\begin{equation}\label{eq:xsConds}
    {\|\nomflow(\mg(t))\|>0} \quad \text{and} \quad \sysstate_\nom(t)\equiv\xs(\mg(t))
\end{equation}
holding for all  $t\in[0,T)$, with $\nomflow(\mg):=\frac{d}{ds}\xs(\mg)$. {Here   $\mg:[0,T)\to{\mgspace}$ is a  homeomorphism, strictly monotonically increasing in time, with its   nominal time evolution over $\mgspace$ is governed by the  autonomous differential equation
\begin{equation}\label{eq:nvelDef}
    \dot{\mg}=\nvel(\mg),    
\end{equation}
 given a known strictly-positive, $\cont{1}$  function $\nvel:\mgspace\to\Rip$ such that $f(\sysstate_\mg(\mg))=\nvel(\mg)\nomflow(\mg)$.
 One can therefore consider the curve parameter $\mg=\mg(t)$ simply as a  rescaling of time along the orbit $\nomorb$, with $\mg= t$ if one takes $\nvel=1$ on $[0,T)$. }
 {There are two main benefits of allowing for such more general parameterizations, for which $\nvel(\mg)=\|f(\xs(\mg))\|/\|\nomflow(\mg)\|\not\equiv 1$, rather than just keeping to the time parameterization: First, it adds some flexibility when planning such motions, as one can fix the interval $\mgspace$, whereas $[0,T)$ will vary depending on the period $T$. For example, in the case of second-order systems, it allows one to specify a path parameterized by $\mg$, and let the traversal velocity along the path be determined by \eqref{eq:nvelDef}. Second, it can be helpful in regards to a so-called \emph{projection operator}---a mapping which can be used to recover the corresponding ``position'' along the orbit, given only knowledge of the system's states within some neighbourhood. We provide the definition of such operators next.}
\begin{definition}\label{def:ProjOp}
    A mapping $\prj:\Ri{n}\supset\prjspace\to\mgspace$  is said to be a \emph{projection operator} onto the curve $\xs:\mgspace\to\nomorb$  if it is $\mathcal{C}^2$  within an open tubular neighbourhood $\prjspace$
     of the orbit $\nomorb$ and it is a left inverse of the curve, that is $\mg\equiv\prj(\xs(\mg))$ for all $\mg\in\mgspace$    .
\end{definition}

In order to also have some measure of the deviation from the orbit, we  will further assume that  a set of $(n-1)$ so-called \emph{transverse coordinates}, denoted by $\tvc=\tvc(\sysstate)$, are {known for the orbit $\nomorb$.}\footnote{Rather than a (minimal) set of $(n-1)$ transverse coordinates, only minor modifications are needed in order to extend the proposed scheme  to also work for an \emph{excessive} number of transverse coordinates; see, e.g., references \citenum{leonov2006generalization,saetre2019excessive,saetre2020excessive}. }
\begin{definition}\label{def:TVC}
    {A $\mathcal{C}^2$, vector-valued function $\tvc:\Ri{n}\to\Ri{n-1}$ is a vector of transverse coordinates for the orbit $\nomorb$ if 
    \begin{equation*}
      \|\tvc(y)\|=0 \ \text{ and } \  \text{rank}\left[\tvcjac(y)\right]=n-1 
    \end{equation*}
    holds  for all $y\in\nomorb$.}
\end{definition}
Note that at least one such  set of  coordinates is guaranteed to exist as the orbit $\nomorb$ is closed \cite{hauser1994converse}.
Moreover, from their definition and the definition of the projection operator $\prj(\cdot)$, it follows that they together constitute a valid change of coordinates, in the sense that the mapping $\sysstate\mapsto (\prj(\sysstate),\tvc(\sysstate))$ is a diffeomorphism in a neighbourhood of the desired orbit. 

{The main value of such a set of transverse coordinates, however, follows from the well-known fact that the exponential stability of the orbit \eqref{eq:NomOrb} is  equivalent to the exponential stability of the origin of their dynamics in a closed-loop system \cite{hauser1994converse,mohammadi2018dynamic}} Therefore, our task will now be to utilize the knowledge of the stabilizing feedback $\nfb(\cdot)$ of the nominal system \eqref{eq:NomNonLinSys} to find a matrix function $S_\perp:\mgspace\to\Ri{m\times n}$ which solves Problem~\ref{prbolem:GenSSforLTP} for the linear-periodic system corresponding to the linearization of the   dynamics of the  coordinates $\tvc$  along the orbit. { We will show next that if such a matrix function can be found, then $\smf(\sysstate):=S_\perp(\prj(\sysstate))\tvc(\sysstate)$ can be taken as a solution to Problem~\ref{prob:MainProblem}. }

%%%%%%%%%%%%%%%%%%%%%%%%
\subsection{Main Result: Transverse coordinates-based switching function design}\label{sec:MR}
{Let $\tvc$ denote a vector of transverse coordinates  by Def.~\ref{def:TVC}. Considering the nonlinear system \eqref{eq:NonLinSys},  their dynamics are given by
\begin{equation}\label{eq:TVD}
    \frac{d}{dt}{\tvc}=f_\perp(\sysstate)+g_\perp(\sysstate)(\ac+\dist(\sysstate,t)),
\end{equation}
where the functions $f_\perp(\sysstate):=\tvcjac(\sysstate)f(\sysstate)$ and $g_\perp(\sysstate):=\tvcjac(\sysstate)g(\sysstate)$ both are derived using the chain rule.}

A key part of our approach utilizes the corresponding \emph{transverse linearization}, which is short for the first-order approximation (linearization) of the transverse dynamics \eqref{eq:TVD} along the nominal solution.\cite{hauser1994converse,shiriaev2008can}

%%%%%%%%%%%%%%%%%%%%%%%%%%%%%%%%%%%%%
\begin{lemma}[Transverse linearization \cite{mohammadi2018dynamic,saetre2020excessive,manchester2011transverse}]\label{lemma:TVL}
 {The linear, $T$-periodic system 
\begin{equation}\label{eq:TransverseLinearization}
    \frac{d}{dt}\delta\tvc=A_\perp(\mg(t)) \delta\tvc+B_\perp(\mg(t))\ac,
\end{equation}
 where  $A_\perp(\mg):=\jac{f}_\perp(\xs(\mg)){(\jac{\tvc})}\pinv(\xs(\mg))$ and $  B_\perp(\mg) :=g_\perp(\xs(\mg))$, corresponds to the linearization of the transverse dynamics \eqref{eq:TVD} when omitting the perturbation (i.e. for $\dist=0$)  along the curve $\xs:\mgspace\to \nomorb$.}
\end{lemma}

{With this in mind, note that the transverse linearization of \eqref{eq:NomNonLinSys} can be written as a differential equation in terms of $\mg$  using \eqref{eq:nvelDef}:
\begin{equation}\label{eq:TVLCL}
    \frac{d }{d \mg} \delta\tvc=\frac{1}{\nvel(\mg)}\left[A_\perp(\mg)+B_\perp(\mg)\lfb_\perp(\mg)\right] \delta\tvc=:\frac{1}{\nvel(\mg)}A_\perp^{cl}(\mg)\delta\tvc
\end{equation}
Here the term involving the matrix function 
$ \lfb_\perp(\mg):=\jac{\nfb}(\xs(\mg)){(\jac{\tvc})}\pinv(\xs(\mg))$ is obtained from the first-order approximation of $\nfb(\cdot)$ in terms of $(\tvc,\prj)$ about a point $\xs(\mg)\in\nomorb$,  using the fact that $\|\nfb(y)\|=\|D\nfb(y)f(y)\|=\|D\nfb(y)\nomflow(\prj(y))\|=0$ for all $y\in\nomorb$, as well as that $(\tvc,\prj)\mapsto h(\tvc,\prj)=\sysstate$ is a diffeomorphism in a neighbourhood of $\nomorb$, with\cite{saetre2020excessive}
\begin{equation*}
    \frac{\partial h}{\partial \tvc}\Bigg\lvert_{\sysstate=\xs(\mg)}=\big(\I{n}-\nomflow(\mg)\jac{\prj}(\xs(\mg))\big)(\jac{\tvc})\pinv(\xs(\mg))\quad  \text{and} \quad \frac{\partial h}{\partial \prj}\Bigg\lvert_{\sysstate=\xs(\mg)}=\nomflow(\mg).
\end{equation*}
} We will denote    by $\stm_\perp^{cl}(\cdot)$ the  state transition matrix corresponding to \eqref{eq:TVLCL}:
\begin{equation*}
    \frac{d}{d\mg}\stm_\perp^{cl}(\mg)=\frac{1}{\nvel(\mg)}A_\perp^{cl}(\mg)\stm_\perp^{cl}(\mg), \quad \stm_\perp^{cl}(0)=\I{n-1}, \quad \mg\in\mgspace. 
\end{equation*}
{As the periodic orbit of \eqref{eq:NomNonLinSys} is (locally) exponentially stable  if, and only if, the origin of transverse linearization is asymptotically stable\cite{hauser1994converse,mohammadi2018dynamic},} it follows that all the $(n-1)$ characteristic multipliers, {i.e. the eigenvalues of the Monodromy matrix}\footnote{Note here that while the state transition matrix with $\mg$-parameterization is  only defined for $\mg\in\mgspace$, one can simply take $\hat{\stm}_\perp^{cl}(t,0):=\stm_\perp^{cl}(\mg(t))$ for $t\in[0.T)$, such that, e.g., $\hat{\stm}_\perp^{cl}(t+kT,0)=\stm_\perp^{cl}(\mg(t))(\mono_\perp^{cl})^k $ for $t\in[0,T)$. }
 $\mono_\perp^{cl}:=\stm_\perp^{cl}(\mg_T)$, necessarily have magnitudes strictly less than one.
 
%%%% Main Result %%%%
This leads us to  the main result of this paper.
\begin{theorem}\label{theorem:mainResult}
    Suppose the state transitions matrix of \eqref{eq:TVLCL} admits a real, $\mg_T$-periodic FL factorization:
    \begin{equation*}
        \stm_\perp^{cl}(\mg)=L(s)e^{\mg F}.
    \end{equation*}
    Further suppose that there exists a full-rank matrix $\hat{S}\in\Ri{
   m\times n}$ such that
   \begin{enumerate}
       \item  $\det [\hat{S}L\inv(\mg)B_\perp(\mg)]\neq 0$,
       \item $\hat{S}z=0$ $\implies$ $\hat{S}Fz=0$,
   \end{enumerate}
   are satisfied for all $\mg\in\mgspace$.
   {  Then, for any projection operator $\prj(\cdot)$ (see Def.~\ref{def:ProjOp}), the  following function  solves Problem~\ref{prob:MainProblem}:
   \begin{equation}\label{eq:sigmaInMainTheorem}
       \smf(\sysstate):={S}_\perp(\prj(\sysstate))\tvc(\sysstate) \quad \text{with} \quad  S_\perp(\mg):=\hat{S}L\inv(\mg).
   \end{equation} 
As a consequence, the desired solution \eqref{eq:xsConds} is  (locally) asymptotically orbitally stable when restricted to the sliding manifold 
\begin{equation*}
\SM:=\left\{\sysstate\in\Ri{n}: \ \smf(\sysstate)=\0{m\times 1}\right\}.    
\end{equation*}
}
\end{theorem}
\noindent 
\textit{Proof.} See Appendix~\ref{proof:theorem:mainResult}.
%%%%%%%%%%%%%%%%%%%%%%%%%%%%%%%%%%%%

\begin{remark}
    The existence of a real $\mg_T$-periodic FL factorization is assumed in Theorem~\ref{theorem:mainResult}  rather than a $2\mg_T$-periodic factorization, which, as we recall, is always guaranteed to exist. This restriction is due to the {image} of the projection operator $\prj(\cdot)$ being equal to $[0,\mg_T)$. More precisely, given a triplet $(L(t),F,Y)$ corresponding to a real, $2\mg_T$-periodic factorization (see e.g. Theorem~\ref{th:realFLtrans} or Theorem 3.1 in  \citenum{montagnier2003real}), we would be limited to only recovering the subinterval $[0,\mg_T)$ through $\prj(\cdot)$, but by our definition of $S_\perp(\mg)$ and considering a factorization as in Theorem~\ref{th:realFLtrans}, we would  naturally require continuity of $S_\perp(\cdot)$ at $\mg=\mg_T$. This then corresponds to the same condition as in Corollary~\ref{corr:SyS}, namely 
    \begin{equation*}
        S_\perp(\mg_T)=\hat{S}Y=S_\perp(0)=\hat{S}. 
    \end{equation*}
    Hence the rows of $\hat{S}$ must then be linear combinations of the eigenvectors of $Y$ corresponding to its unitary eigenvalues, which is trivially true whenever $Y=\I{n-1}$, i.e. when one has a $\mg_T$-periodic factorization.
\end{remark}

\begin{remark}
    The existence of an FL factorization of course does not in turn imply the existence of a (unique) real invariant subspace of $F$ satisfying the conditions of the Theorem; see the discussion after Lemma~\ref{lemma:SM_LTI}.
\end{remark}

\begin{remark}
{
In the special cases 
when the dynamical system \eqref{eq:NonLinSys} is so-called
\emph{transversely feedback linearizable}\cite{banaszuk1995feedback,nielsen2008local},
then one can instead simply utilize the theory outlined in Section~\ref{sec.LTI}, in particular Lemma~\ref{lemma:SM_LTI}, in order to find a solution to Problem~\ref{prob:MainProblem}. More specifically, since  there then  exist (at least locally) transverse coordinates and a smooth feedback transformation $\ac=a(\sysstate)+b(\sysstate)v$, $v\in\Ri{m}$, such that the transverse dynamics \eqref{eq:TVD} can be written as $\frac{d}{dt}\tvc=A\tvc+B\big(v+(b(\sysstate))\inv\dist(\sysstate,t)\big)$, where the pair $(A,B)$ is controllable, the statements in Section~\ref{sec.LTI} are evidently readily applicable. 
}
\end{remark}
%%%%%%%%%%%%%%%%%%%%%%%%
%%%% Control Design %%%%
\subsection{Some comments regarding  stabilization of the sliding manifold }\label{sec:ContDes}
{Although it is the design of  sliding manifolds which is the main focus of this paper, we will will in this section briefly demonstrate how the presented scheme allows for adding a robustifying feedback extension to an existing orbitally stabilizing feedback. The following statement may act as a useful stepping stone towards the design of such extensions.}
%%%%%%%%%%%%%%%
\begin{lemma}\label{lemma:PreContDesign}
     For some projection operator  $\mg=\prj(\sysstate)$, let
     \begin{equation*}
        \sigma(\sysstate):=S_\perp(\prj(\sysstate))\tvc(\sysstate)=\hat{S}L\inv(\prj(\sysstate))\tvc(\sysstate)  
     \end{equation*}
     be a switching function according to Theorem~\ref{theorem:mainResult}. If   the controller in \eqref{eq:NonLinSys} is taken as
    \begin{equation}\label{eq:PreSMcont}
    u=k(\sysstate)+  v, \quad v\in\Ri{m},
    \end{equation}
    then the dynamics of $\sigma(\sysstate)$ outside of the sliding manifold $\SM$  are given by
    \begin{equation}\label{eq:sigmaDot}
    \dot{\sigma}=\mathcal{F}_\sigma \sigma +S_\perp(\mg)\big[B_\perp(\mg)+\Tilde{B}_\perp(\tvc,\mg)\big]\big( v+\dist(\sysstate,t)\big)+R_\sigma(\tvc,\mg) \quad \text{with} \quad \mg=\prj(\sysstate),
\end{equation}
 and where  $\mathcal{F}_\sigma:=\hat{S}F\hat{S}\pinv\in\Ri{m\times m}$ is Hurwitz,    while $\|\tilde{B}_\perp(\tvc,\mg)\|=\bigO(\|\tvc\|)$ and $\|R_\sigma(\tvc,\mg)\|=\bigO(\|\tvc\|^2)$ for all $\mg\in\mgspace$.
\end{lemma}
%%%%%%%%
\noindent 
\textit{Proof.} See Appendix~\ref{proof:lemma:PreContDesign}.
\begin{remark}
    The known nominal  feedback $k(\sysstate)$ does not have to be included in \eqref{eq:PreSMcont} due to the equivalent control \eqref{eq:EqCont} when confined to the sliding manifold. However, its inclusion  has two clear benefits: 1) it allows reducing the magnitude of the  gains used in the (discontinuous) extension $v$,  which may help to alleviate chattering; and 2) it can increase the rate of convergence to the sliding manifold, especially when the system states are far away from it.
\end{remark}
%%%% Control design %%%% 
The next step is then to design a feedback extension $v\in\Ri{m}$ in \eqref{eq:PreSMcont} such that the sliding manifold $\Sigma:= \{\sysstate\in\prjspace: \ \smf(\sysstate):={S}_\perp(\prj(\sysstate)) \tvc(\sysstate)\equiv 0 \}$ is reached in finite time. In a similar manner to the control law proposed in Proposition~\ref{prop:LinCL} for the LTI system, we will  suggest  for this purpose a unit-vector approach: 
\begin{equation}\label{eq:SMCu}
    v={\begin{cases}-\zeta(\sysstate)\left(S_\perp(\prj(\sysstate))B_\perp(\prj(\sysstate))\right)\inv\frac{\sigma(\sysstate)}{\|\sigma(\sysstate)\|}, \quad &\text{if} \quad \sigma \neq0,  \\
        \quad \0{m\times 1} \quad &\text{if} \quad \sigma = 0 ,
        \end{cases}}  
\end{equation}
where $\zeta:\Ri{n}\to\Rip{}$ is $\mathcal{C}^1$. While there is no general guarantee for the feasibility such a control scheme due to the nonlinearity of the problem, we  derive some conditions upon the gain $\zeta$ and the nonlinear system in general in the following.
     
  Since $\mathcal{F}_\sigma$ is Hurwitz, denote by $P=P\transp\in\Ri{m\times m}$  the  unique positive definite (PD) solution to $\mathcal{F}_\sigma\transp P+P\mathcal{F}_\sigma=-2Q$ for some symmetric PD matrix $Q\in\Ri{m\times m}$, and consider the Lyapunov function candidate $V_\sigma:=2^{-1}\sigma\transp(\sysstate)P\sigma(\sysstate)$. We have
    \begin{align*}
        \dot{V}_\sigma&=2^{-1}\sigma\transp\left[\mathcal{F}_\sigma\transp P+P\mathcal{F}_\sigma-2\frac{\zeta}{\|\sigma\|}P\right]\sigma
        +\sigma\transp P W
        \\
        &\le-\left(\lambda_{\min}(Q)\|\sigma\|+\lambda_{\min}(P)\zeta-\lambda_{\max}(P)\|W\|\right)\|\sigma\|,
    \end{align*}
    where $W:=-S_\perp\tilde{B}_\perp(S_\perp B_\perp)\inv\zeta \frac{\sigma}{\|\sigma\|}+S_\perp(B_\perp+\tilde{B}_\perp)\dist+R_\sigma$. Since $S_\perp B_\perp$ is nonsingular for all $\mg\in\mgspace$, there exist (known) positive constants $c_0,\hat{c}_0>0$ such that
        $\|S_\perp(\mg)B_\perp(\mg)\|\le c_0$ and $\|(S_\perp(\mg)B_\perp(\mg))\inv\|\le \hat{c}_0$.
    Furthermore, we will assume that a pair of smooth class $\mathcal{K}$  functions (see Def.~4.2 in \citenum{khalil2002nonlinear}), denoted by $c_1,c_2:\Rip{}\to\Rip{}$, are known  such that for all $\mg\in\mgspace$:
    \begin{equation}
        \|S_\perp(\mg)\tilde{B}_\perp(\tvc,\mg)\|\le c_1(\|\tvc\|) \quad \text{and} \quad \|R_\smf(\tvc,\mg)\|\le c_2(\|\tvc\|).
    \end{equation}
    Then $W$  has the following known upper bound:
    % \begin{equation*}
        $\|W\|\le \zeta \hat{c}_0 c_1(\|\tvc\|)+(c_0+c_1(\|\tvc\|))\dist_M+c_2(\|\tvc\|).$
    % \end{equation*}
    We may therefore ensure that $\dot{V}_\sigma$ is (locally) negative definite by taking $\zeta$ as to satisfy 
    \begin{equation*}
        \zeta >\frac{\lambda_{\max}(P)(c_0+c_1(\|\tvc\|))\dist_M+\lambda_{\max}(P)c_2(\|\tvc\|)-\lambda_{\min}(Q)\|\sigma\|}{\lambda_{\min}(P)-\lambda_{\max}(P)\hat{c}_0 c_1(\|\tvc\|)}.
    \end{equation*}
    For some $0<\upsilon\ll 1$, this is always possible  within 
    \begin{equation}\label{eq:SMCtube}
        \mathcal{N}(\upsilon):=\left\{\sysstate\in\prjspace: \|\tvc(\sysstate)\|\le c_1\inv\left(\frac{\lambda_{\min}(P)(1-\upsilon)}{\lambda_{\max}(P)\hat{c}_0}\right)\right\},
    \end{equation}
    as then $0<\upsilon\lambda_{\min}(P)\le\lambda_{\min}(P)-\lambda_{\max}(P)\hat{c}_0 c_1(\|\tvc\|) $ for all $\tvc\in\mathcal{N}(\upsilon)$.
    Indeed,  taking
    \begin{equation}\label{eq:SMCzetaNonlinear}
        \zeta(\sysstate)=\frac{\sqrt{\lambda_{\max}(P)}}{{\upsilon\lambda_{\min}(P)}}\left[\mu_\star+{\sqrt{\lambda_{\max}(P)}}\left((c_0+c_1(\|\tvc(\sysstate)\|))\dist_M+c_2(\|\tvc(\sysstate)\|)\right)\right]
    \end{equation}
     ensures that $\dot{V}_\sigma<-\mu_\star\sqrt{V_\sigma}-V_\smf\lambda_{min}(Q)/\lambda_{max}(P)$ for all $\tvc\in\mathcal{N}(\upsilon)\backslash\{\0{} \}$. 
     
     It is important to note that this alone does not guarantee that the sliding manifold will be reached, as the system states may escape $\mathcal{N}(\upsilon)$ beforehand.  This can  be resolved through additional assumptions, for example requiring  local input-to-state stability of  the transverse dynamics 
     \begin{equation}\label{eq:TVDCL}
        \frac{d}{dt}{\tvc}=f_\perp(\sysstate)+g_\perp(\sysstate)(\nfb(\sysstate)+v+\dist(\sysstate,t))
     \end{equation}
     with respect to the input $(v+\dist)$, for $v$ taken  within a certain admissible range.
     
     For the sake of brevity, we will here simply assume the forward invariance  (see Def.~\ref{def:ForwardInvariance} in the Appendix)   of $\mathcal{N}(\upsilon)$ with respect to \eqref{eq:TVDCL}. The above may then summarize as follows.
        
\begin{proposition}\label{prop:ContDesign}
     Let the conditions in Lemma~\ref{lemma:PreContDesign} be satisfied and consider \eqref{eq:TVDCL}  with the feedback extension \eqref{eq:SMCu}. Suppose
      that by taking $\zeta:\Ri{n-1}\to\Rip$ satisfying \eqref{eq:SMCzetaNonlinear} for some $\upsilon\in (0,1)$, the  tube $\mathcal{N}(\upsilon)$ is forward invariant with respect to the transverse dynamics \eqref{eq:TVDCL}. Then any solution of the nonlinear system \eqref{eq:NonLinSys} starting inside $\mathcal{N}(\upsilon)$ reaches the sliding manifold \eqref{eq:SM} in finite time.
\end{proposition}

%%%%%%%%%%%%
%%%% LRC %%%
\subsection{Comparison to a Lyapunov redesign controller}\label{sec:LRC}
Suppose the symmetric matrix function ${R}_\perp:\mgspace\to\Ri{(n-1)\times (n-1)}$ is the unique PD solution to the  periodic Lyapunov equation 
\begin{equation}
    {\nvel(\mg)\frac{d}{d\mg}{{R}}_\perp(\mg)}+ A_\perp^{cl}(\mg)\transp R_\perp(\mg)+{R}_\perp(\mg) A_\perp^{cl}(\mg)=-{Q}_\perp(\mg)
  \end{equation}
for some symmetric, continuous, PD matrix function  ${Q}_\perp:\mgspace\to\Ri{(n-1)\times( n-1)}$, where we have used 
{$\dot{{R}}(\mg(t))=\nvel(\mg(t))\frac{d}{d\mg}R(\mg(t))$}.   Then by defining    $\xi(\mg,\tvc):=B_\perp\transp(\mg)R_\perp(\mg)\tvc$,  the following \emph{Lyapunov redesign controller} \cite{khalil2002nonlinear} (LRC) provides an alternative robustifying feedback extension:
\begin{equation}\label{eq:UnitVectorRobstExt}\    \ac=\nfb(\sysstate)-\zeta(\sysstate)\Xi(\prj    (\sysstate),\tvc), \quad              \Xi(\mg,\tvc):=
    \begin{cases} \frac{\xi(\mg,\tvc)}{\|\xi(\mg,\tvc)\|} \quad &\text{if} \quad  \|\xi(\mg,\tvc)\| \neq 0,  \\
    \quad \0{m\times 1} \quad &\text{if} \quad  \|\xi(\mg,\tvc)\| = 0.
\end{cases} 
\end{equation}
Here   $\zeta:\Ri{n}\to\Rip{}$ is a   smooth function which must be taken sufficient large as to dominate the disturbance term $\dist$, while $\prj(\cdot)$ is some projection operator. 

The  inspiration for this controller comes from references \citenum{gutman1979uncertain} and  \citenum{corless1981continuous}, and is based on the fact that $\hat{V}_\perp=\sysstate_\perp\transp R_\perp(\prj(\sysstate))\sysstate_\perp$ will be a  Lyapunov function for the  orbit of the nominal nonlinear system   \eqref{eq:NomNonLinSys}. Hence for $\zeta$ taken sufficiently large, it guarantees local negative definiteness of the time  derivative of $\hat{V}_\perp$ despite of any matched disturbances. 

 There are some key behavioral differences between the LRC  \eqref{eq:UnitVectorRobstExt} and the proposed SMC. For instance,   while the SMC \eqref{eq:SMCu} is designed as to reach the sliding manifold and render it invariant, the sole purpose of the LRC \eqref{eq:UnitVectorRobstExt} is to ensure the local negative definiteness of the derivative of the Lyapunov function $\hat{V}_\perp$.   One may therefore expect the LRC to locally ensure strict contraction of the Lyapunov function candidate at the expense of having little prior knowledge of its convergence rate to the target obit. The proposed SMC, on the other hand, will locally ensure a specific convergence rate (depending on the magnitude of the characteristic exponents of the chosen invariant subspaces)  when in sliding mode, but with little control over the contraction towards the target orbit when in the reaching phase.
 
%%%%%%%%%%%%%%%%%%%%%%%    
%%%% Cart-pendulum %%%%
%%%%%%%%%%%%%%%%%%%%%%%
\section{Case Study: Oscillation Control of the Cart-Pendulum System}\label{sec:CPmainSec}
In order to  construct a switching surface of the form \eqref{eq:sigmaInMainTheorem}  utilizing the method outlined in Section~\ref{sec:NonLinSysTVC}, the following four  basic ingredients first have to be obtained:
\begin{itemize}
    \item[1)] A desired {periodic solution} of  a nominal model of the system,  parameterized on the form \eqref{eq:xsConds};
    \item[2)] A {projection operator}  recovering the parameterizing variable of this solution (see Def.~\ref{def:ProjOp});
    \item[3)] A set of transverse coordinates for the solution (see Def.~\ref{def:TVC}); and lastly,
    \item[4)] An {exponentially} orbitally stabilizing state feedback $k(\cdot)$ for the nominal (disturbance-free) system.
\end{itemize}
In order to both demonstrate how these ingredients can be obtained and then used in the synthesis of a robust orbitally stabilizing feedback, we will in this section consider the concrete example of oscillation control of the well-known cart-pendulum system. 
We will utilize the virtual  constraints approach of Reference \citenum{shiriaev2005constructive} for both trajectory generation and orbital stabilization for the nominal system.

Note that Appendix~\ref{sec:UndAcMechSys} contains a brief outline of this approach, together with some related statements which may be used to obtain these ingredients for mechanical (Euler-Lagrange) systems with  $n_q$ degrees of freedom and one degree of underactuation.

\begin{figure}
	\centering
	\includegraphics[]{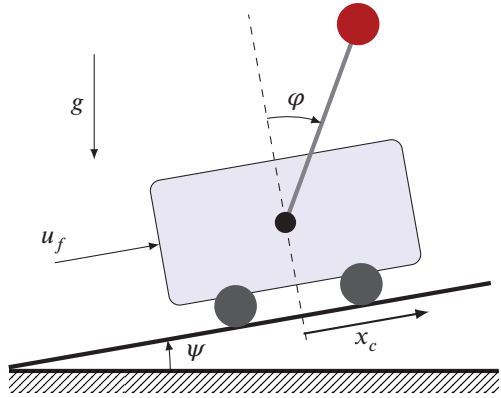}
	\caption{Schematic of the cart-pendulum system.}
	\label{fig:cart_pendulum EL}
\end{figure}
%%%%%%%%%%%%%%%%%%%%%%%%%%
\subsection{System model}
The system consists of an unactuated pendulum attached to a cart. The cart is situated on a ramp of constant inclination $\psi$ and is driven by an external force $u_f$. The schematic of the system and the coordinate convention used is shown in Figure~\ref{fig:cart_pendulum EL}.

We will differentiate between the ``real'' (actual) model of the system dynamics, which we do not know exactly, and a nominal (approximation)  model which we will use to construct a desired periodic solution and to design a nominal feedback.
\paragraph{Real dynamical model:}
The real equations of motion of the system are given by 
\begin{subequations}\label{eq:CPeqTrue}
\begin{align}\label{eq:CPeqTrue1}
    (m_c+m_p)\ddot{x}_c+m_pl_p\cos(\varphi)\ddot{\varphi}-m_pl_p\sin(\varphi)\dot{\varphi}^2 
    +g(m_c+m_p)\sin(\psi)&={u}_f-\upsilon_c\text{sign}(\dot{x}_c)+d_x(t),\\     \label{eq:CPeqTrue2}
    (m_pl_p^2+J_p)\ddot{\varphi}+m_p l_p\cos(\varphi)\ddot{x}_c
    -m_pl_pg\sin(\varphi-\psi)&=-\upsilon_p\text{sign}(\dot{\varphi})+d_p(t).
\end{align}
\end{subequations}
Here $g=\SI{9.81}{\meter\per\second\squared} $ is the gravitational acceleration, $m_c$ and $m_p$ denote the mass of the cart and pendulum bob, respectively; $l_p$ is the length of the pendulum and $J_p$ is its moment of inertia; $\upsilon_c$ and $\upsilon_p$ are dry friction coefficients; while $d_x(t)$ and $d_p(t)$ are smooth,  bounded, time-varying disturbances.

\paragraph{Nominal (disturbance-free) model used for trajectory generation and in the nominal feedback synthesis:} 
The following assumptions are made for the  nominal system: zero inclination of the ramp ($\psi=0$), no friction ($\upsilon_c=\upsilon_p=0$),  unit masses ($m_c=m_p=1$), the  pendulum is as a point mass of unit length ($l_p=1$ and $J_p=0$) and zero disturbances ($d_x=d_p=0$).

The equations of motion for the nominal system are then described by 
\begin{subequations}\label{eq:CPnom}
\begin{align}\label{eq:CPeq1}
    2\ddot{x}_c+\cos(\varphi)\ddot{\varphi}-\sin(\varphi)\dot{\varphi}^2&=u_f,\\
    \ddot{\varphi}+\cos(\varphi)\ddot{x}_c-g\sin(\varphi)&=0.\label{eq:CPeq2}
\end{align}
\end{subequations}

%%%%%%%%%%%%%%%% 
\subsection{Trajectory generation}
Using the nominal model of the cart-pendulum system \eqref{eq:CPnom}, our goal is now to plan a periodic trajectory which corresponds to oscillations of pendulum about its up-right position.  As in Reference \citenum{shiriaev2005constructive} (see also Appendix~\ref{sec:UndAcMechSys}), we will for this purpose look for a solution along which, for $\gc:=[x_c;\varphi]$, one has $\gc^\nom(t)\equiv \vc(\mgvc^\nom(t))$ for some function  $\mgvc^\star(t)$ and where
\begin{equation}\label{eq:CPvc}
    \vc(\mgvc):=\begin{bmatrix}-a\sin(\mgvc) \\ \mgvc \end{bmatrix}, \quad  a\in\Ri{}.
\end{equation}
 That is, a solution along which $\mgvc^\nom(t)\equiv \varphi^\nom(t)$ and  $x_c^\nom(t)+a\sin(\varphi^\nom(t))\equiv 0$ hold for all $t\in\Rip$.

Assuming the invariance of $\gc=\vc(\mgvc)$, such that $\Dgc=\vc'(\mgvc)\dot{\mgvc}$ and $\DDgc=\vc'(\mgvc)\ddot{\mgvc}+\vc''(\mgvc)\dot{\mgvc}^2$, with $\vc'=\frac{d}{d\mgvc}\vc$ and where we have dropped the $\nom$-superscript for readability, we may rewrite \eqref{eq:CPnom} as
\begin{subequations}
\begin{align}
    \label{eq:CPnomCont}
    (1-2a)\cos(\mgvc)\ddot{\mgvc}+(2a-1)\sin(\mgvc)\dot{\mgvc}^2&=\ac_f\\
    \label{eq:CPredDyn}
    (1-a\cos^2(\mgvc))\ddot{\mgvc}+a\cos(\mgvc)\sin(\mgvc)\dot{\mgvc}^2-g\sin(\mgvc)&=0.
    \end{align}
\end{subequations}
It follows that $\gc^\nom(t)\equiv \vc(\mgvc^\nom(t))$ is a solution to \eqref{eq:CPnom} only if $\mgvc^\nom(t)$ is a solution to \eqref{eq:CPredDyn}, with the corresponding nominal control input  $\ac_f^\nom(t)$ given by \eqref{eq:CPnomCont}  {with $\ddot\theta$ substituted from \eqref{eq:CPredDyn}.}
As we are interested in oscillations about the up-right equilibrium of the pendulum, which corresponds to $(\mgvc,\dot{\mgvc})=(0,0)$, we will utilize the fact that this is an equilibrium point  of type center for \eqref{eq:CPredDyn} provided that $a>1$ (see Proposition 1 in \citenum{shiriaev2005constructive}).  Moreover, we will later also utilize the fact that the function $I=I(\mgvc,\dot{\mgvc},\mgvc_0,\dot{\mgvc}_0)$, defined by
\begin{equation}\label{eq:CPintFunc}
    {I=\frac{1}{2}(1-a\cos^2(\mgvc))\left[(1-a\cos^2(\mgvc))\dot{\mgvc}^2-(1-a\cos^2(\mgvc_0))\dot{\mgvc}_0^2 +2g(\cos(\mgvc)-\cos(\mgvc_0))\right],}
\end{equation}
is zero along any solution of \eqref{eq:CPredDyn} with initial conditions $(\mgvc_0,\dot{\mgvc}_0)$  (see, e.g., Proposition~\ref{prop:intFunc} in the Appendix). 

%%%%%%%%%%%%%%%%%%%
\subsection{Choosing a parameterization and  projection operator}
Suppose that a $T$-periodic solution $\mgvc_\star(t)=\mgvc_\star(t+T)$ of \eqref{eq:CPredDyn}  has been found which {encircles} its (center) equilibrium point $(0,0)$.
The next step is then to obtain a regular parameterization and a projection operator\footnote{Choosing a  projection operator is not strictly necessary already at this this stage, but it can  be convenient to do so while simultaneously choosing the parameterization. }. {In this regard, first note that  we cannot use $\mgvc$ to parameterize the curve as $\dot{\mgvc}_\nom(t)$ will not  be strictly positive everywhere along the orbit.}
Thus  we instead note that the time derivative of $\mg_\nom(t)=\text{atan2}(-\dot{\mgvc}_\nom(t),\mgvc_\nom(t))$, with $\text{atan2}(\cdot)$ denoting the four-quadrant arctangent function, is given by
\begin{equation*}
    \frac{d}{dt}\dot{\mg}_\nom(t)=\frac{\dot{\mgvc}_\nom^2(t)-{\mgvc}_\nom(t)\ddot{\mgvc}_\nom(t)}{{\mgvc}_\nom^2(t)+\dot{\mgvc}_\nom^2(t)}.
\end{equation*}
{It follows that if $\dot{\mg}_\nom(t)>0$ for all $t\in[0,T)$, then we can take $\nvel:[0,2\pi)\to\Rip$  satisfying $\dot{\mg}_\nom(t)\equiv\nvel(\mg_\nom(t))$.} Moreover,  {abusing notation by  considering $\theta_\nom(\cdot)$ and $\dot{\theta}_\nom(\cdot)$ as functions of $\mg$,} it allows us to use the parameterization $\mg\mapsto (\theta_\nom(\mg),\dot{\theta}_\nom(\mg))$ such that 
\begin{equation}\label{eq:CPparamAndProjOp}
    \xs(\mg)=\begin{bmatrix}\vc(\mgvc_\nom(\mg))\\ \vc'(\mgvc_\nom(\mg))\dot{\mgvc}_\nom(\mg)\end{bmatrix} \quad\text{and}\quad \prj(\sysstate)=\text{atan2}(-\dot{\varphi},\varphi)
\end{equation}
correspond, respectively, to a regular parameterization \eqref{eq:xsConds} and a projection operator (see Def.~\ref{def:ProjOp}) for the target motion.

%%%%%%%%%%%%%%%%%%%%%%%%%%%
\subsection{Feedback transformation and transverse coordinates}
Given a periodic solution to \eqref{eq:CPredDyn} with  initial condition $(\mgvc_0,\dot{\mgvc}_0)$,  we have the following candidates for transverse coordinates:
\begin{equation}
\label{eq:CPtransverseCoordinates}    
\tvc=\begin{bmatrix}y \\ \dot{y}\\ I \end{bmatrix}:=\begin{bmatrix}
 x_c+a\sin(\varphi) \\ \dot{x}_c+a\cos(\varphi)\dot{\varphi} \\
 {\frac{1}{2}\alpha(\varphi)\left[\alpha(\varphi)\dot{\varphi}^2-\alpha(\varphi_0)\dot{\varphi}_0^2 +2g(\cos(\varphi)-\cos(\varphi_0))\right]}
\end{bmatrix},
\end{equation}
{where $\alpha(\varphi):=1-a\cos^2(\varphi)$. }
Indeed, by \eqref{eq:CPintFunc}, one has $\|\tvc(\xs(\mg))\|\equiv 0 $ for all $\mg\in\mgspace$, while both $\tvc$ and its Jacobian matrix are locally well defined if $a\cos^2(\theta_\nom(t))\neq 1$ for all $t\in[0,T)$.

In order to linearize the dynamics of these coordinates, we first introduce the feedback transformation
\begin{align}\label{eq:CPfeedbackTransform}
    \ac_f= \frac{\sin(\varphi)(2a-1)}{1-a\cos^2(\varphi)}\left(\dot{\varphi}^2-g\cos(\varphi) \right)
     +\frac{1+\sin^2(\varphi)}{1-a\cos^2(\varphi)}\ac, \quad \ac\in\Ri{}.
\end{align}
This transformation is partially feedback linearizing, in the sense that it results in $\ddot{y}=u$.\footnote{Notince from $\ddot{y}=\ac$ the possibility of pre-stabilizing the $(y,\dot{y})$-subspace by taking $u=\hat{u}-\hat{k}_y y-\hat{k}_{\dot{y}}\dot{y}$ for any constant gains $\hat{k}_{{y}},\hat{k}_{\dot{y}}>0$.} Thus the transverse dynamics can be written as {${\frac{d}{dt}{\tvc}}=\hat{A}_\perp(\varphi(t),\dot{\varphi}(t))\tvc+\hat{B}_\perp(\varphi(t),\dot{\varphi}(t))\ac$,} where
\begin{align*}
    \hat{A}_\perp(\varphi,\dot{\varphi}):=
    \begin{bmatrix}0 \quad ~ & 1 & 0  \\
        0 \quad ~ & 0 & 0 \\
        0 \quad ~ & 0 & \dot{\varphi}\frac{2a\cos(\varphi)\sin(\varphi)}{1-a\cos^2(\varphi)}\end{bmatrix} \quad \text{and} \quad
    \hat{B}_\perp(\varphi,\dot{\varphi}):=\begin{bmatrix} 0 \\ 1 \\ -\dot{\varphi}(1-a\cos^2(\varphi))\cos(\varphi )
    \end{bmatrix}.
\end{align*}
The transverse linearization (see \eqref{eq:TransverseLinearization}) may then be obtained by simply using the previously found parameterization $\mg\mapsto(\mgvc_\nom(\mg),\dot{\mgvc}_\nom(\mg))$, that is: $A_\perp(\mg)=\hat{A}_\perp(\mgvc_\nom(\mg),\dot{\mgvc}_\nom(\mg))$ and $B_\perp(\mg)=\hat{B}_\perp(\mgvc_\nom(\mg),\dot{\mgvc}_\nom(\mg))$.

%%%%%%%%%%%%%%%%%%%%%%%%%%%%%%%%%%%%
\subsection{Designing an Orbitally Stabilizing Feedback  for the Nominal System}\label{sec:LQR}
The design of a nominal feedback, corresponding to $\nfb(\cdot)$ in Problem~\ref{prob:MainProblem}, is then last  piece of the puzzle which is required before we can apply the proposed sliding mode control synthesis. For this purpose, we will utilize the following well-known statement (see, e.g., Reference \citenum{yakubovich1975}).
\begin{proposition}\label{prop:PRDE}
    Suppose there exists 
    a symmetric, positive definite (SPD) matrix function $R_\perp:[0,\mg_T)\to\Ri{(n-1)\times(n-1)}$ which for all $\mg\in[0,\mg_T)$ is the solution to the periodic Riccati Differential equation (PRDE)
    \begin{align} \label{eq:PRDE}
        &\nvel(\mg)\frac{d}{d\mg}R_\perp(\mg)+A_\perp\transp(\mg)R_\perp(\mg)+R_\perp(\mg)A_\perp(\mg)+Q(\mg)
        -R_\perp(\mg)B_\perp(\mg)\Gamma^{-1}(\mg)B_\perp\transp(\mg)R_\perp(\mg)=0
    \end{align}
    given smooth SPD matrix functions $Q(\mg)\in\Ri{(n-1)\times(n-1)} $ and $\Gamma(s)\in\Ri{m\times m}$. Then the origin of the closed-loop system
    \begin{equation*}
        {\frac{d}{dt}\delta\tvc=\big[A_\perp(\mg(t))-B_\perp(\mg(t))\Gamma^{-1}(\mg(t))B_\perp\transp(\mg(t))R_\perp(\mg(t))\big]\delta\tvc}
    \end{equation*}
    corresponding to taking $\ac=\lfb_\perp(\mg)\delta\tvc$ in \eqref{eq:TransverseLinearization} with
    \begin{equation}
        \lfb_\perp(\mg):=-\Gamma^{-1}(\mg)B_\perp\transp(\mg)R_\perp(\mg),  \label{eq:LQRfeedback}%
    \end{equation}
    is exponentially  stable.
\end{proposition}
    
It  follows that if $\lfb_\perp(\cdot)$ is taken according to \eqref{eq:LQRfeedback}, then taking $\ac=\nfb(\sysstate)=\lfb_\perp(\prj(\sysstate))\tvc$ in \eqref{eq:NonLinSys} for some projection operator $\prj(\cdot)$  renders the desired orbit \eqref{eq:NomOrb} an exponentially stable limit cycle of the disturbance-free system \eqref{eq:NomNonLinSys}.

%%%%%%%%%%%%%%%%%%%
\subsection{Constructing a Switching Function for the Cart-Pendulum system}
%%%%%%%%%%%%%%%%%%%
\begin{figure}
    \centering
        \begin{minipage}{.45\textwidth}
    \includegraphics[width=1.1\linewidth]{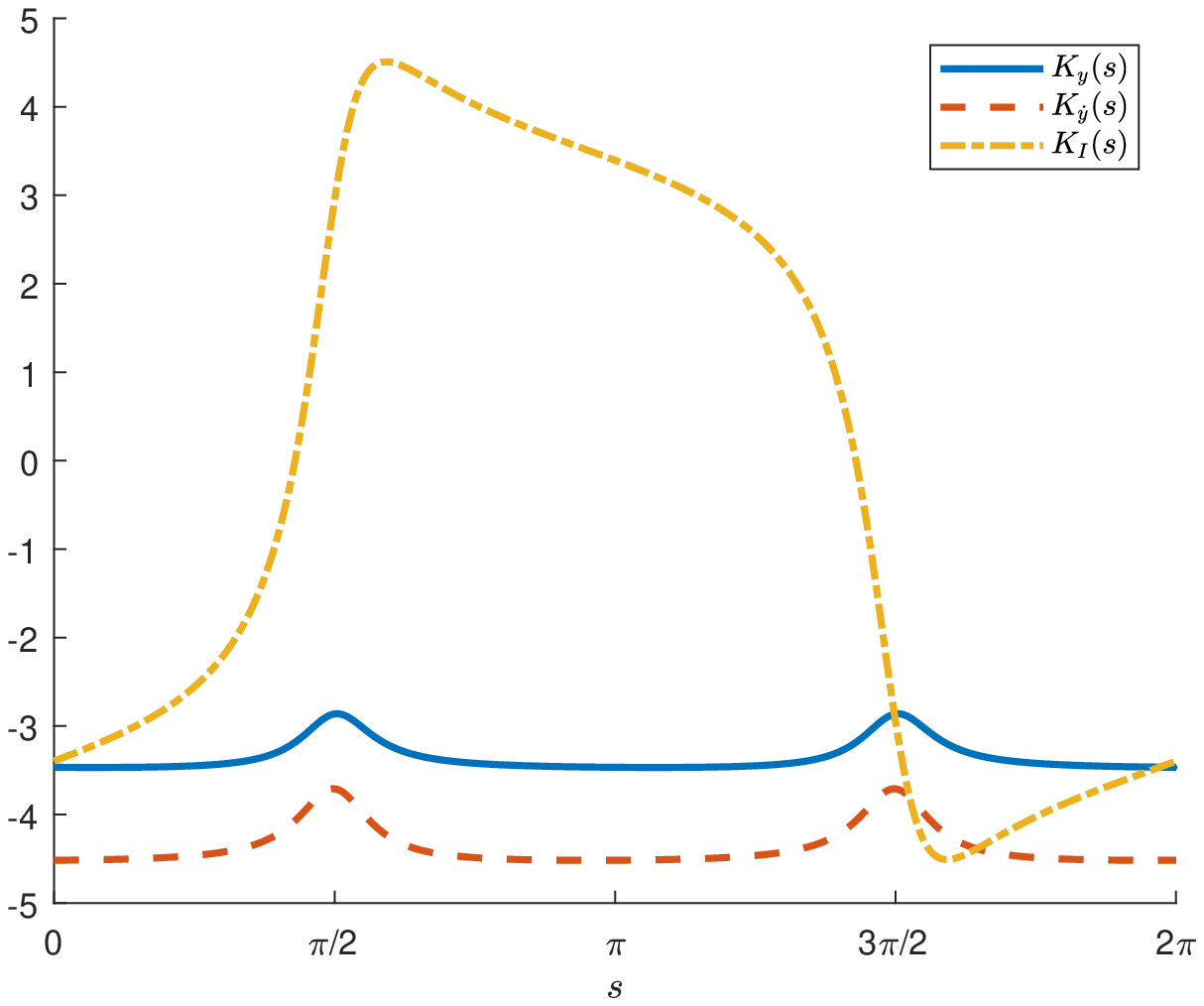}
    \caption{Elements of the  nominal feedback matrix $\lfb_\perp(\mg)$ given by  \eqref{eq:LQRfeedback} found by solving the PRDE \eqref{eq:PRDE}.}
    \label{fig:Kfeedback}
    \end{minipage}
    \qquad
    \begin{minipage}{.45\textwidth}
    \centering
    \includegraphics[width=1.1\linewidth]{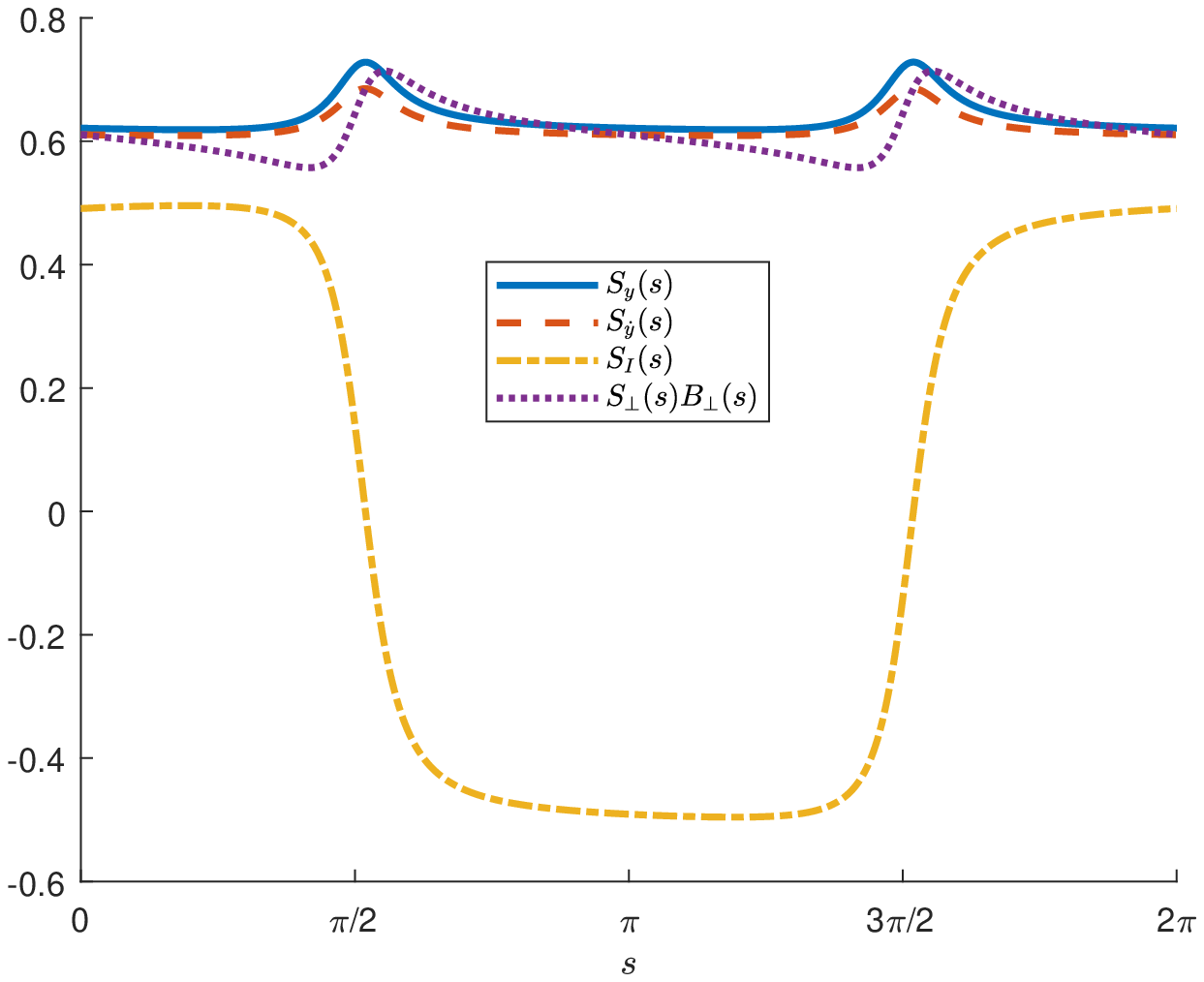}
    \caption{Elements of the designed switching function $S_\perp$ given by \eqref{eq:sigmaInMainTheorem}. Here   $(S_\perp B_\perp)$ is seen to be nonsingular. }
    \label{fig:switchingFunction}
    \end{minipage}
\end{figure}
%%%%%%%%%%%%%%%%%%
 In this subsection, we will now design a switching function of the form \eqref{eq:sigmaInMainTheorem} for the cart-pendulum system. 
 We will consider the periodic orbit corresponding to \eqref{eq:CPvc} with $\mgvc_\nom(t)=\mgvc_\nom(t+T)$ the solution to \eqref{eq:CPredDyn} for
 \begin{equation}
    a=1.5 \quad \text{and} \quad (\mgvc_0,\dot{\mgvc}_0)=(0,0.5).
 \end{equation}
 Using \eqref{eq:CPintFunc}, one can  find that the amplitude of the induced oscillations of the pendulum is approximately $\SI{0.113}{\radian}$, such that the transverse coordinates are  well defined within a neighbourhood of the nominal orbit ($a\cos^2(\varphi)> 1$ for $|\varphi|< \sqrt{(1/a)}\approx \SI{0.62}{\radian}$). 
 
 Using the transverse coordinates \eqref{eq:CPtransverseCoordinates} and  the feedback transformation  \eqref{eq:CPfeedbackTransform}, as well as using the parameterization and projection operator given by \eqref{eq:CPparamAndProjOp} (note that these are then also locally well defined), a  nominal LQR feedback controller was designed for the system using Proposition~\ref{prop:PRDE}: The PRDE \eqref{eq:PRDE} with $Q=\I{3}$ and $\Gamma=0.1$  was solved using the method proposed in Reference \citenum{gusev2016sdp}, in which $R_\perp(\mg)$ was approximated by a truncated Fourier series of order 100 for 500 evenly spaced sampling points. The elements of the matrix $K_\perp(\mg)=[K_y(\mg),K_{\dot{y}}(\mg),K_I(\mg)]$ found from \eqref{eq:LQRfeedback} are shown in Figure~\ref{fig:Kfeedback}.

In order to construct a switching function of the form \eqref{eq:sigmaInMainTheorem}, we therefore need to find: 1)  a real $2\pi$-periodic  FL factorization of the state-transition matrix of \eqref{eq:TVLCL}, and then 2) a full-rank left-annihilator of a real invariant subspace of $F$ of co-dimension $m$.\footnote{Since the system is mechanical, the dimension of the transverse dynamics will always be odd, that is $(n-1)=2n_q-1$. Thus by Statement~\ref{statement:invaraintSSodd} in Appendix~\ref{app:GenIS}, the system \eqref{eq:TVLCL} must then have at least one such subspace. Although  there is no guarantee that this subspace has an annihilator such that $S_\perp(\mg)B_\perp(\mg)$ is nonsingular everywhere.} 

To find an FL factorization ($L(\mg),F,Y)$ for \eqref{eq:TVLCL}, we used the boundary value problem formulation proposed in \citenum{montagnier2004control}.\footnote{An  initial condition for $F$ was found by integrating \eqref{eq:STM} and using  \eqref{eq:realPLF} in the Appendix.} This resulted in $Y=\I{3}$, i.e. $L(\tau)$ was $2\pi$-periodic as required, while the found  matrix $F$ was approximately given by
\begin{equation*}
    F \approx
    \begin{bmatrix}
    0.0843  &  1.1269 &   0.8987 \\
   -3.4618  & -4.4920 &  -3.1910 \\
    0.4531  &  0.4799 &  -0.0735
    \end{bmatrix}.
\end{equation*}
The three real eigenvalue-eigenvector pairs of this $F$ are in turn approximately given by 
\begin{equation*}
    (\lambda_1,\upsilon_1)\approx\Big(-2.97,\begin{bmatrix}-0.32  \\   0.94 \\  -0.11\end{bmatrix}\Big), \quad (\lambda_2,\upsilon_2)\approx\Big(-1.06,\begin{bmatrix} 0.68 \\  -0.73   \\ 0.04\end{bmatrix}\Big), \quad (\lambda_3,\upsilon_3)\approx\Big(-0.45,\begin{bmatrix}0.02  \\ -0.63  \\  0.78\end{bmatrix}\Big).
\end{equation*}
The matrix $F$ therefore has three real invariant subspaces of co-dimension one:
\begin{equation*}
    \text{span}\{\upsilon_1,\upsilon_2\}, \quad \text{span}\{\upsilon_1,\upsilon_3\} \quad \text{and} \quad \text{span}\{\upsilon_2,\upsilon_3\}.     
\end{equation*}
However, only the subspace $\text{span}\{\upsilon_2,\upsilon_3\}$ had an annihilator that  satisfied the nonzero-determinant condition 
in Theorem~\ref{theorem:mainResult}. We therefore took $\hat{S}$ such that $\hat{S}\upsilon_2=\hat{S}\upsilon_3\equiv 0$ (i.e. $\hat{S}\approx[0.62,0.61,0.48]$). The corresponding  matrix function   $S_\perp(\mg)=\hat{S}L\inv(\mg)=[S_y,S_{\dot{y}},S_I]$  is shown in Figure~\ref{fig:switchingFunction}, in which $\big(S_\perp(\mg)B_\perp(\mg)\big)$ can be seen to be separated  from zero.

Note here that    $\text{span}\{\upsilon_2,\upsilon_3\}$ consists of the two one-dimensional subspaces whose eigenvalues have the smallest magnitude. Since solutions of the nominal (disturbance-free) system under just the LQR feedback may also partly correspond to the subspace spanned by $\upsilon_1$, it is therefore to be expected  that the convergence close to the orbit  will be somewhat slower in general when confined to the induced sliding manifold.

%%%%%%%%%%%%%%%%%%%%%%%

\begin{table}
    \centering
    \caption{Parameters of the cart-pendulum system \eqref{eq:CPeqTrue} used in simulation for the three considered scenarios.  }
    \label{tab:SystemParams}
    {\small
    \begin{tabular}{l| c c c c c c c c c }
    \toprule
         &   $m_c$ & $m_p$ & $l_p$ & $J_p$ & $\psi$& $\upsilon_c$ & $\upsilon_p$  & $d_x(t)$ & $d_p(t)$ 
         \\
         Types of perturbations
         &   [\SI{}{\kilo\gram}] & [\SI{}{\kilo\gram}] & [\SI{}{\meter}] & [\SI{}{\kilo\gram \meter\squared}] & [\SI{}{\radian}] & [\SI{}{\newton}] & [\SI{}{\newton \meter}] &  [\SI{}{\newton}]& [\SI{}{\newton \meter}] 
         \\
          \midrule
        None (nominal case) & 1 & 1& 1 & 0 & 0 & 0 & 0 & 0 & 0      
        \\ 
         Only matched & 1 & 1& 1 & 0 & 0 & 0.25 & 0 & $0.1 \sin(t)$ & 0
         \\ 
         Matched \& unmacthed &  1.2 & 1.2 & 0.9 & 0.2 & $5\pi/180$ & 0.25 & 0.1 & $0.1 \sin(t)$ & $0.1 \sin(t)$
         \\ \bottomrule
    \end{tabular}
    }
\end{table}
  
%%%%%%%%%%%%%%%%%%%%%%%%%%%
\subsection{Simulation results}\label{sec:CPsim}
Using the above designed switching function, we will now compare in simulations  the control law \eqref{eq:PreSMcont} with feedback extension \eqref{eq:SMCu}  to both the nominal LQR and the Lyapunov redesign controller (LRC) given by \eqref{eq:UnitVectorRobstExt}. 
These were implemented as follows:
\begin{align*}
    \tag{LQR}\label{eq:simContLQR}
      \ac_{LQR}&=\lfb_\perp(\mg) \tvc,  &\mg&=\prj(\sysstate),
      \\
      \tag{SMC}\label{eq:simContSMC}
      \ac_{SMC}&=\lfb_\perp(\mg) \tvc-{\mu_1}\ \sat\left({\sigma(\mg,\tvc)}/{\epsilon}\right),  &\sigma(\mg,\tvc)&:=S_\perp(\mg)\tvc,
      \\ 
    \tag{LRC}\label{eq:simContrLQR}
      \ac_{LRC}&=\lfb_\perp(\mg)\tvc-\mu_2\ \sat\left(\xi(\mg,\tvc)/{\epsilon}\right), &\xi(\mg,\tvc)&:={B_\perp\transp(\mg) R_\perp(\mg)\tvc}.
\end{align*}
Here the saturation function $\sat(\cdot)$ was used with  $\epsilon=10^{-3}$ rather than the signum function in order to mitigate chattering \cite{khalil2002nonlinear}. The  gains $\mu_1,\mu_2>0$ in the SMC and LRC were taken as constants to facilitate the comparison between the controllers, as  well as to highlight the effects of these gains with respect to the magnitude of any matched disturbances. 

%%%% Nominal %%%%
\begin{figure}[ht!]
\centering
\includegraphics[width=\linewidth]{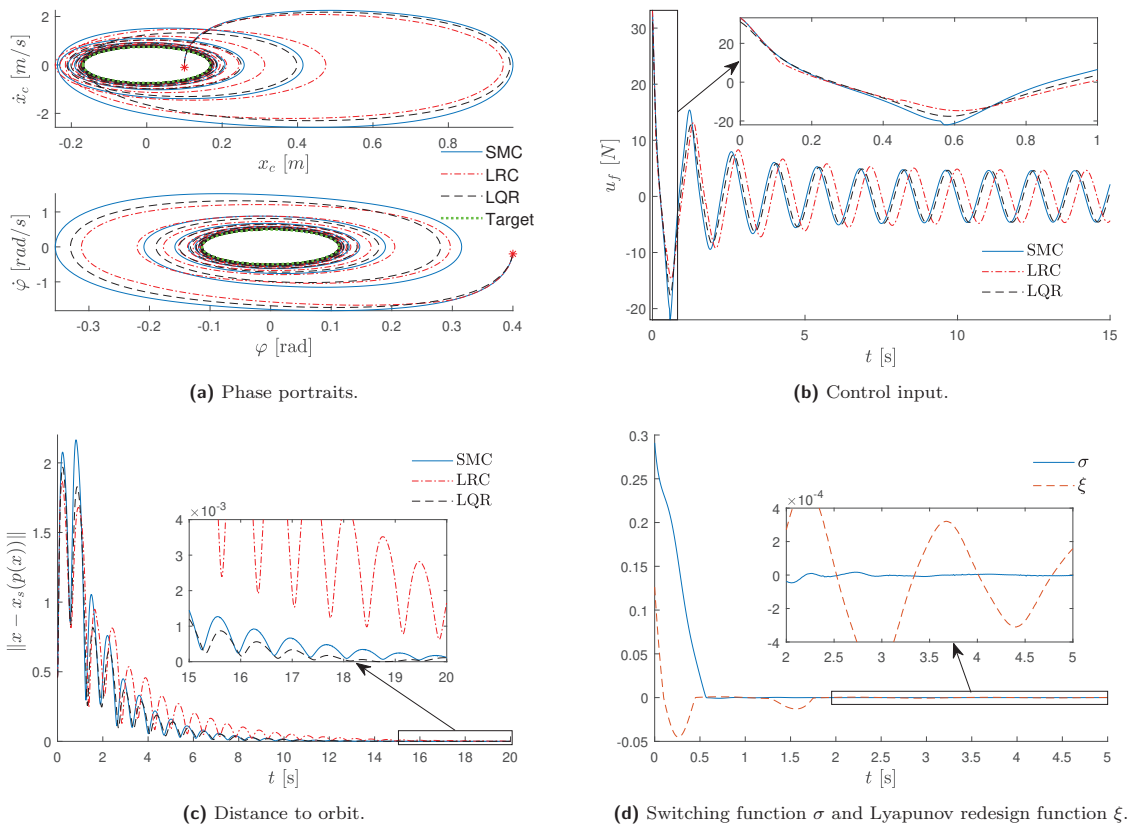}
    \caption{(Nominal model) Simulation results showing the performance of each controller, with $(\mu_1,\mu_2)=(0.5,0.5)$,  when implemented on the  disturbance-free system;
    SMC ({\color{blue} ----}), LRC ({\color{red} $-\cdot-$}), LQR({\color{black} $-\ -$}),  Target orbit ({\color{green} $\cdots$}).  }
    \label{fig:None}
\end{figure}
%%%%%%%%%%%%%%%%%%%%%%%%%%
Each controller was tested on the system \eqref{eq:CPeqTrue} for  three different scenarios: without any perturbations (corresponding to the nominal system \eqref{eq:CPnom}), with only matched perturbations, and with both matched and unmatched perturbations. Table~\ref{tab:SystemParams} contains the parameters used in the dynamic model  for each of these scenarios. The initial conditions were taken as 
\begin{equation*}
 (x_c(0),\varphi(0),\dot{x}_c(0),\dot{\varphi}(0))=(0.1,0.4,-0.1,-0.2).
\end{equation*}

%%%% Nominal %%%%
Figure~\ref{fig:None} shows the simulation results when implemented on the nominal system \eqref{eq:CPnom}, with the gains of the SMC and LRC taken as    $(\mu_1,\mu_2)=(0.5,0.5)$. Both the convergence to the orbit and the control inputs are seen to be fairly similar for all the control laws. As seen in (d), however, the LRC  quickly drives the states close to the manifold $\xi\equiv 0$ such that $u_{LRC}$ remains close to zero. This results in  a slightly slower convergence to the target orbit for the LRC than the other two controllers. The SMC, on the other hand, can be seen in (c) to have a {larger overshoot than the other controllers with respect to the projection operator--based distance measure $\|\sysstate-\xs(\prj(\sysstate))\|$}, before eventually having a similar convergence rate to that of the nominal LQR after reaching the sliding manifold.

\begin{figure}[ht!]
\centering
\includegraphics[width=1\linewidth]{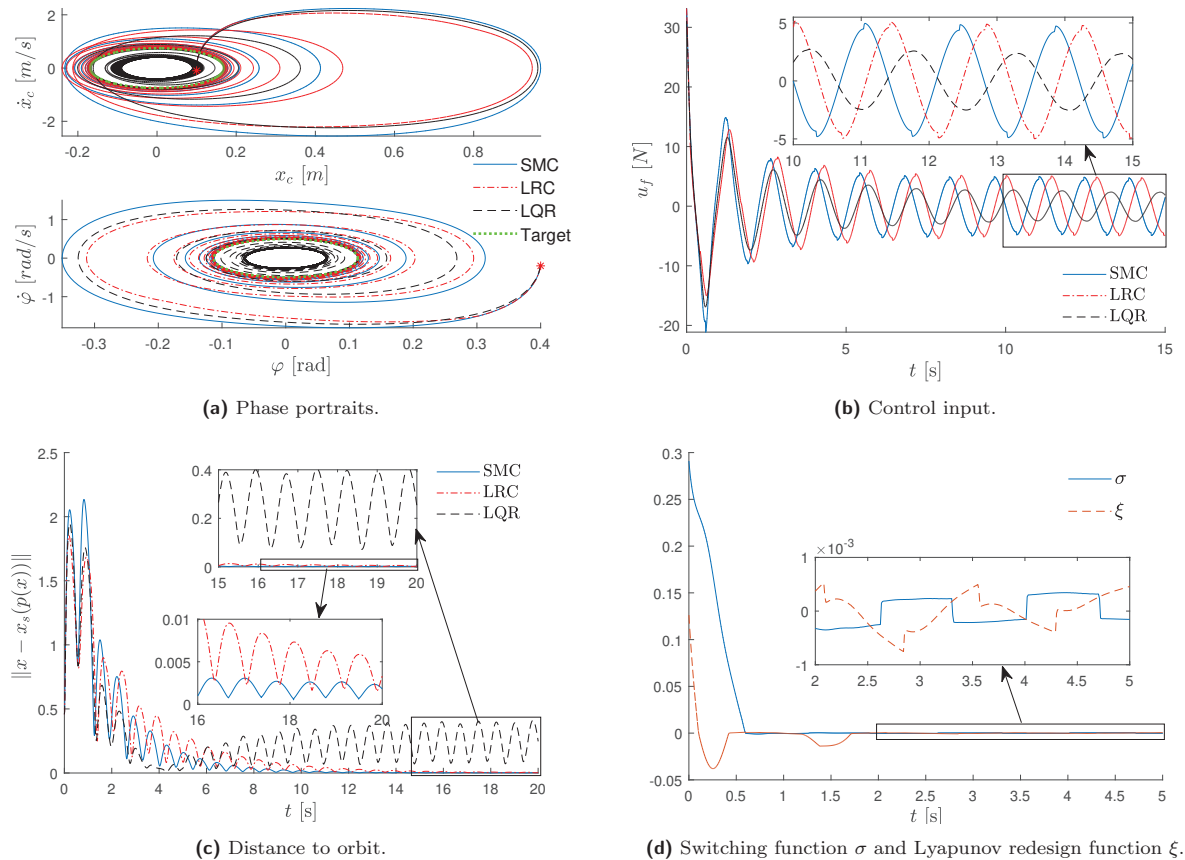}
    \caption{(Matched perturbations) Simulation results showing the performance of each controller, with $(\mu_1,\mu_2)=(0.5,0.5)$,  when subject to only matched perturbations;
    SMC ({\color{blue} ----}), LRC ({\color{red} $-\cdot-$}), LQR({\color{black} $-\ -$}), Target orbit ({\color{green} $\cdots$}).  }
    \label{fig:Matched}
\end{figure}

\begin{figure}[ht!]
\centering
\includegraphics[width=1\linewidth]{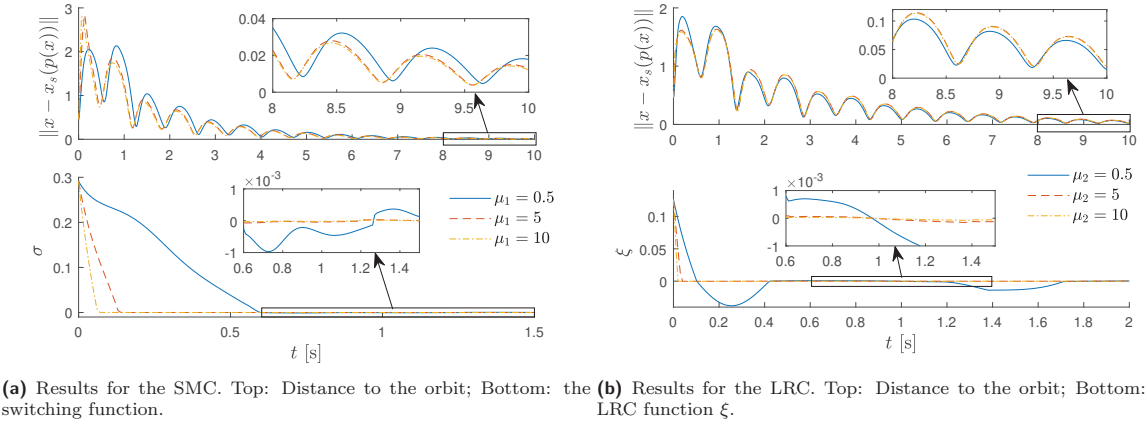}
    \caption{Gain comparison for the SMC (\textbf{a}) and LRC (\textbf{b}) when subject to  matched disturbances; $\mu_i=0.5$ ({\color{blue} ----}), $\mu_i=5$ ({\color{red} $-\cdot-$}), $\mu_i=10$ ({\color{yellow} $-\ -$}).}
    \label{fig:gainsMacthed}
\end{figure}

%%%% Matched %%%%
Figure~\ref{fig:Matched} shows the performance of the three controllers when subject to only matched perturbations. The perturbation term had the upper bound $\Delta_M\le 0.35$ (see Table~\ref{tab:SystemParams}) and the gains of the SMC and LRC were again taken as  $(\mu_1,\mu_2)=(0.5,0.5)$. As seen in (a), the LQR is unable to ensure convergence to the target orbit and instead settles into a perturbed orbit having a lower amplitude, whereas both the SMC and LRC are able to almost completely reject the disturbances. The effects of increasing the gains of both controllers can be seen in Figure~\ref{fig:gainsMacthed}. As one would expect, increasing the gain for the SMC is seen to decrease the time needed to reach the sliding manifold, slightly speeding up the convergence, but at the expense of both larger {overshoot} and increased peak actuator forces. The opposite behaviour may be observed when increasing the gains of the LRC;  it can be seen that the system's states are driven faster towards the manifold $\xi\equiv 0$, leading to slightly slower  convergence to the target orbit, but reducing the { overshoot}.

%%%% Gain comparison %%%%

\begin{figure}[ht!]
\centering
\includegraphics[width=1\linewidth]{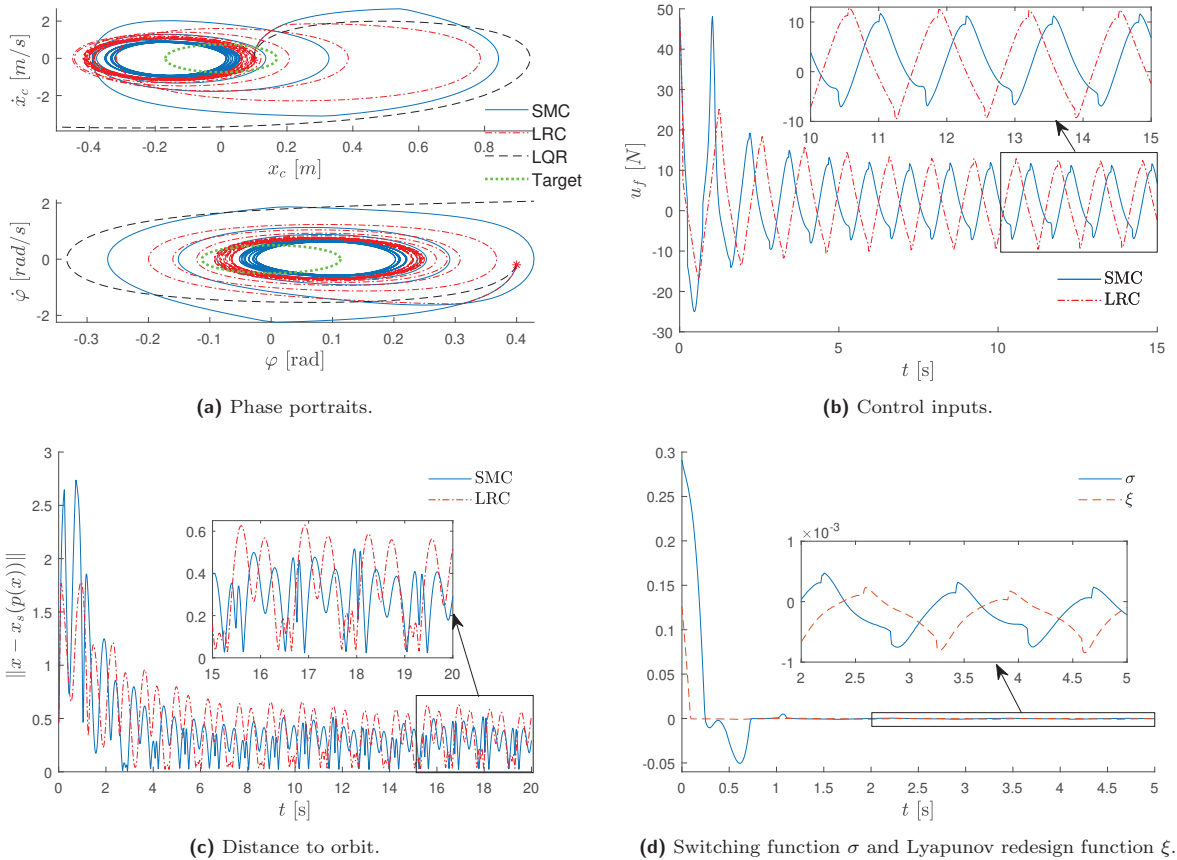}
    \caption{(Matched \& unmacthed  perturbations) Simulation results showing the performance of each controller, with $(\mu_1,\mu_2)=(4,4)$,  when subject to both matched and unmatched disturbances;
    SMC ({\color{blue} ----}), LRC ({\color{red} $-\cdot-$}), LQR({\color{black} $-\ -$}), Target orbit ({\color{green} $\cdots$}).  }
    \label{fig:Unmatched}
\end{figure}

%%%% Unmacthed %%%% 
Figure~\ref{fig:Unmatched} shows the response of the system under the three controllers for $(\mu_1,\mu_2)=(4,4)$ when subject to both matched and unmatched perturbations. The system under the LQR is seen to become unstable as the controller was unable to keep the angle of the pendulum within the region where the feedback \eqref{eq:CPfeedbackTransform} and the third transverse coordinate are well defined, corresponding to $|\varphi|< \sqrt{(1/a)}\approx \SI{0.62}{\radian}$. To stay within this region, both the gain of the SMC and LRC had to be increased, with the SMC needing the largest increase for the considered initial conditions. It can again be seen from (b) and (c) in Figure~\ref{fig:Unmatched}  that the SMC is initially  more aggressive than the LRC, leading to high peaks in the force applied to the cart and to larger {a larger  overshoot with respect to the target orbit.} After the transient phase, however, the SMC settles into an orbit  that is closer to the nominal orbit on average and which requires slightly less control forces than the settling orbit of the LRC.

\begin{figure}[ht!]
\centering
\includegraphics[width=1\linewidth]{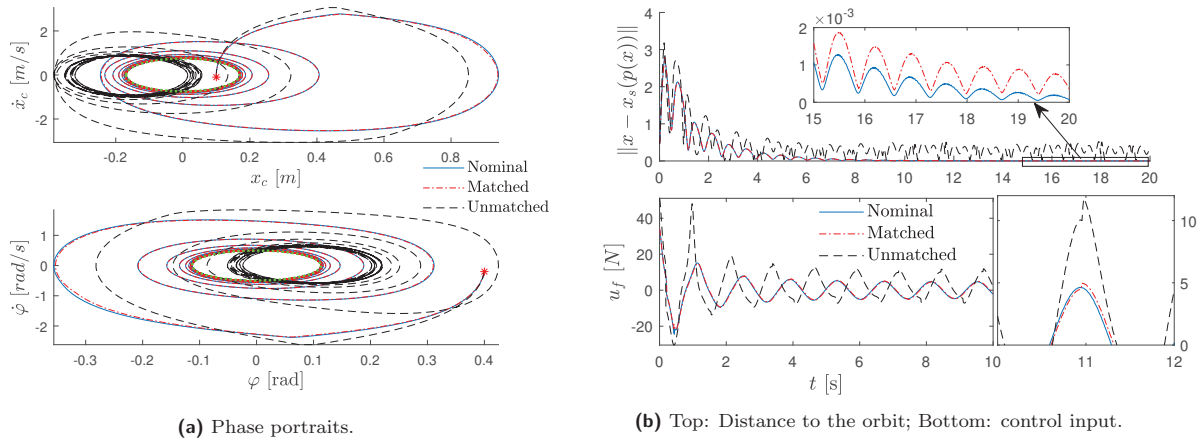}
    \caption{System response with only the sliding mode controller. The gain was taken as $\mu_1=2$ for both the nominal case ({\color{blue} ----}) and with only matched disturbances  ({\color{red} $-\cdot-$}), while $\mu_1=6$ when subject to both matched and unmatched disturbances ({\color{black} $-\ -$}).}
    \label{fig:SMConly}
\end{figure}

Figure~\ref{fig:SMConly}  shows the results for the different  scenarios  when taking in \eqref{eq:CPfeedbackTransform}  the pure sliding mode controller given by
\begin{equation*}
    u=-\mu_1\text{sat}(\sigma(\mg,\tvc)/\epsilon).
\end{equation*} 
  The gain had to be increased to maintain a similar performance in all three scenarios:  $\mu_1=2$ for both the nominal case and with only matched perturbations, while it was increased to $\mu_1=6$ for the case also including unmatched perturbations. The similarity of these results compared to those under the controller \eqref{eq:simContSMC} indicates that the equivalent control when confined to the manifold indeed corresponds to that of design controller \eqref{eq:simContLQR} used in the switching function synthesis.
  
\begin{figure}[ht!]
\centering
\includegraphics[width=1\linewidth]{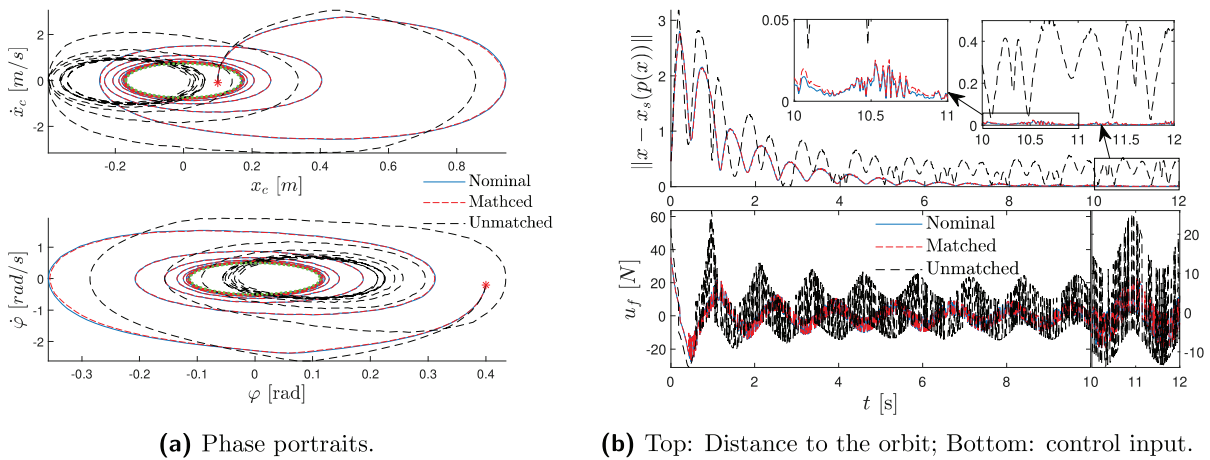}
    \caption{The same scenarios as in Figure~\ref{fig:SMConly} but with white noise added to the measurements passed to the controllers.}
    \label{fig:SMConlyWithNoise}
\end{figure}

  {In order to test the sensitivity of the control scheme to measurements noise, we added a small amount of white noise (signal-to-noise ratio of \SI{50}{\decibel}) to the state measurements passed to the controller for the same scenarios as in Figure~\ref{fig:SMConly}. The obtained responses are shown in Figure~\ref{fig:SMConlyWithNoise}.  It can be seen that, while the measurement noise leads to  chattering in the control signals if $\epsilon$ is not increased, it has little effect upon the overall response in all three scenarios.}

%%%% Only SMC %%%%

%%%% Only SMC with noise %%%%
%%%%%%%%%%%%%%%%%%%%
%%%% Conclusion %%%%
\section{Concluding Remarks and Future Directions}\label{sec:conclusion}
The task of robustifying a known orbitally stabilizing feedback controller was considered in this paper. For this purpose, a new constructive procedure for generating a switching function was proposed,  allowing for the use of sliding mode control extensions for disturbance rejection. The designed switching function corresponded to an annihilator of a real invariant subspace of the Monodromy matrix of the first-approximation (linearization) of a nominal model of the system. It was constructed using a real Floquet--Lyapunov transformation of state-transition matrix of the linearized dynamics of a set of transverse coordinates along the nominal orbit. This design was complemented with a suggestion for a  unit-vector based approach for stabilizing the corresponding sliding manifold in finite time. 

The feasibility and advantages of the proposed scheme was demonstrated using simulations by considering the challenging task of stabilizing oscillations about the up-right equilibrium of the Cart-Pendulum system subject to both matched- and unmatched perturbations. The proposed sliding mode controller (SMC) was compared to a Lyapunov redesign controller (LRC)  constructed using the knowledge of a Lyapunov function candidate of the nominal system. Both controllers {were shown on a numerical example} to reject matched perturbations and to also handle certain unmacthed perturbations provided the gains were taken sufficiently large. The simulation results demonstrated that whereas the SMC approach had a faster rate of convergence toward the nominal orbit than the LRC {during the transient}, it might  have larger deviations away from it compared to the LRC. 

The contributions of this paper lays the foundation for further research in several directions, {including the following.}

    \emph{Numerical construction of the switching function:} At presented, the proposed scheme rests on two major assumptions: 1) the existence of a real Floquet--Lyapunov factorization of minimal periodicity; and 2) the existence of a real invariant subspace of the Monodromy matrix satisfying certain conditions. Thus the design of numerically tractable  solutions for both evaluating their existence and for their construction are important next steps.

    \emph{Alternative sliding mode controllers:} A  unit-vector controller was suggested for the robustifying feedback extension. In practice, a continuous approximation of this controller must be utilized in order to mitigate chattering. However, such an approximation can only ensure convergence to a boundary layer of the sliding manifold. Exploring alternative continuous sliding mode controllers, such as variants of the super-twisting algorithm \cite{nagesh2014multivariable,lopez2019generalised}, is therefore of interest.
    
      \emph{Generic design using an  excessive set of transverse coordinates:} The suggested scheme requires knowledge of a set of transverse coordinates. While such coordinates may always be found (using, for example, the  virtual constraints approach for mechanical systems), their construction will often require additional numerical steps. Extending the approach of this paper  to also allow for the  use of a more generic \emph{excessive} set of transverse coordinates may therefore be of some value, as such coordinates can be computed knowing just a regular parameterization of the orbit and a projection operator that defines a Moving Poincaré section; see, e.g.,  references \citenum{saetre2020excessive} and \citenum{leonov2006generalization}.

     \emph{Extension to hybrid systems and to non-periodic motions: } The concept of orbital stabilization is  not strictly limited to only periodic motions. For instance, it may be applied to certain finite-time trajectories (e.g. point-to-point motions) or to the  quasi-periodic motions that arise in  hybrid dynamical system (e.g. walking gaits of bipedal robots). The method proposed in this paper may therefore be used for such tasks as well, provided that the transverse linearization along the motion is real reducible.  

%%%%%%%%%%%%%%%%%%%%%%%%%
%%%% {Bibliography} %%%%%
\bibliography{references}%

%%%%%%%%%%%%%%%%%%%%%%%%%
%%%%%% Appendix %%%%%%%%%
\appendix
 \setcounter{equation}{0}
\renewcommand{\theequation}{\thesection\arabic{equation}}
%%%%%%%%%%%%%%%%%%%%%%%%%
\section{Supplementary Material}\label{app:SupMat}
%%%%%%%%%%%%%%%%%%%%%%%%%
%%% Stable manifolds %%%%
\subsection{On constructing stable invariant manifolds from the first-order approximation}\label{app:InvariantManifolds}
We first provide a definition.
\begin{definition}\label{def:ForwardInvariance}
    Let  $h:\Ri{n}\to \Ri{n}$ be $\cont{2}$. A set $\Lambda\subset\Ri{n}$ is said to be   \emph{forward invariant} with respect to
    \begin{equation}\label{eq:simpleDynSys}
        \dot{y}=h(y), \quad y\in\Ri{n}, \quad t\in\Ri{+},
    \end{equation}
    if for any solution $y(\cdot)$ of \eqref{eq:simpleDynSys} satisfying $y(t_0)\in \Lambda$ for some $t_0\in\Ri{+} $,  $y(t)\in \Lambda$ for all $t\ge t_0$.  
\end{definition}

In the particular case when $h(\cdot)$ is a linear map of the form $h(y)=Hy$ for some $H\in\Ri{n\times n}$, then it is well known that $\Lambda^d\subset \Ri{n}$ is a real invariant, $d$-dimensional subspace of \eqref{eq:simpleDynSys} if, and only if,  $H\Lambda^d\subseteq \Lambda^d$ and $\Lambda^d$ is spanned by $d$ real, linearly-independent vectors $v_1,\dots,v_d\in\Ri{n}$.\footnote{Note that a brief outline  of  how one can construct such subspaces using the matrix's real Jordan form  is provided in Appendix~\ref{app:GenIS}. } 

By the Hartman--Grobman theorem, this may also be used to locally approximate the stable invariant manifolds of nonlinear systems of the form \eqref{eq:simpleDynSys}. Indeed, suppose $h(0)=0$ such that \eqref{eq:simpleDynSys} may be written as 
\begin{equation}\label{eq:InvaraitnSubpsacesThmHartman}
    \dot{y}=\hat{H}y+\hat{h}(y), 
\end{equation}
where $\hat{H}:=\jac{{h}}(0)$ has no eigenvalues {on the imaginary axis} and  the $\cont{1}$-mapping $\hat{h}:\Ri{n}\to \Ri{n}$ satisfies $\|\hat{h}(y)\|=\bigO(\|y\|^2)$.
Then any (real) $d$-dimensional, exponentially stable invariant subspace of $\hat{H}$ implies the existence of an exponentially   stable (locally)  invariant manifold of the same dimension for the nonlinear system \eqref{eq:simpleDynSys} about its origin \cite{hartman2002ordinary}.

This fact is also useful in regards to designing switching functions whose zero-level set has desired properties.
For instance, suppose that $\Lambda^{(n-m)}\subset\Ri{n}$ is a real,  stable invariant subspace  of $\hat{H}$ of co-dimension $m (<n)$;  that is  $\hat{H}y\in\Lambda^{(n-m)}$ for all $y\in\Lambda^{(n-m)}$, or equivalently $\hat{H}\Lambda^{(n-m)}\subseteq\Lambda^{(n-m)}$. Let  the vectors  $\upsilon_1,\upsilon_2,\dots,\upsilon_{(n-m)}\in\Ri{n}$ form a basis of $\Lambda^{(n-m)}$. At the same time, these vectors must also span the $(n-m)$-dimensional nullspace of some full rank matrix $S\in\Ri{m\times n}$, that is $S\upsilon_i=0$ for all $i\in\{1,\dots,(n-m)\}$. Or in other words: there exists a  matrix $S\in\Ri{m\times n}$ of full rank such that $Sy=\0{m\times 1}$ if, and only if, $y\in\Lambda^{(n-m)}$. 

Suppose, therefore, that the right-hand side of \eqref{eq:simpleDynSys} is complemented by the term $B(u+\Delta)$ where $u\in\Ri{m}$ are controls, $\dist\in\Ri{m}$ a perturbation and    $B\in\Ri{n\times m}$ is such that $SB\in\Ri{m\times m}$ is nonsingular. Further suppose that a feedback $u$ can be designed such that, despite the higher-order terms, it brings the system's states   onto and renders invariant the sliding manifold $\{y\in\Ri{n}: \ \sigma(y):=Sy\equiv 0\}$. 
Since this manifold corresponds to a  stable invariant subspace of $\hat{H}$, there must consequently exists a nonzero neighbourhood of the origin in which the states exponentially convergence towards it regardless of the disturbance, and despite $\hat{H}$ possibly also having eigenvalues with positive real parts.

%%%%%%%%%%%%%%%%%%%%%%%%%
%%% Invariant subspaces
\subsection{On constructing real invariant subspaces for  LTI systems}\label{app:GenIS}
Consider the following task:
     For any matrix $A\in\Ri{n\times n}$, find all its real invariant subspaces of dimension $d<n$. That is, find any subspace $\Lambda^d\subset\Ri{n}$ spanned by $d$ linear independent vectors $v_1,v_2,\dots,v_d\in \Ri{n}$ such that $Ax\in\Lambda^d$ for all $x\in\Lambda^d$.

It is well known that any (possibly complex) invariant subspace of a matrix $A\in\Ri{n\times n}$ is spanned by its generalized eigenspaces \cite{teschl2012ordinary}. For example, given a real eigenvalue $\lambda^r$ of $A$, any vector in the eigenspace $\mathcal{E}_{\lambda^r}:=\text{ker}\left(\lambda^r\I{n}-A\right)$ spans a real, one-dimensional invariant subspace of $A$, while if $\text{dim}(\mathcal{E}_{\lambda^r})=d>1$, then the basis vectors of $\mathcal{E}_{\lambda^r}$ can be used to generate real invariant subspaces of all dimensions up to and including $d$. 

More generally, one can utilize the fact that any real, square matrix has a real Jordan form   \cite[Thm. 3.4.1.5]{horn2012matrix}:  there exists a nonsingular matrix $V\in\Ri{n\times n}$ and a block diagonal matrix $J\in\Ri{n\times n}$ such that $AV=VJ$.  In this regard, let $J^r_1,\dots,J_{k_r}^{r}$ 
denote the blocks of $J$ corresponding to the real eigenvalues of $A$, and let $V^r_i=[v_{i,1}^r,\dots, v_{i,k_i^r}^r]\in\Ri{n\times k_i^r}$ denote the corresponding columns of $V$ such that  
$ AV^r_i=V^r_iJ_i^r$.
Note that this is equivalent to a Jordan chain:
\begin{align*}
   \left(A-\I{n}\lambda_i^r\right)v_{i,1}^r=0,  \qquad
   \left(A-\I{n}\lambda_i^r\right)v_{i,2}^r=v_{i,1}^r,
   \qquad \dots \qquad
   \left(A-\I{n}\lambda_i^r\right)v_{i,k_i^r}^r=v_{i,(k_i^r-1)}^r.
\end{align*}
Thus for any positive integer $\mu\le k_i^r$,  one may construct a real invariant, $\mu$-dimensional subspace of $A$  spanned by the real, linearly independent generalized eigenvectors $v_{i,1}^r,v_{i,2}^r,\dots,v_{i,\mu}^r$. Furthermore, given two different such generalized eigenspaces, denoted $\{v_{i,1}^r,v_{i,2}^r,\dots,v_{i,k_i^r}^r\}$ and $\{v_{j,1}^r,v_{j,2}^r,\dots,v_{j,k_j^r}^r\}$,  one can construct invariant subspaces of any dimension less than or equal to $k_r^i+k_r^j$; for example, $\Lambda^3=\vspan \{v_{i,1}^r,v_{j,1}^r,v_{j,2}^r\} $, with $i\neq j$,  would be a three-dimensional invariant subspace, and so on.

For the  complex conjugate eigenvalue pairs  of $A$,  denoted $\{\lambda_i^c, \overline{\lambda_i^c}\}$, this, however, cannot be applied directly  as the corresponding generalized eigenspaces, that is
$\mathcal{E}_{\lambda_i^c}^n:=\text{ker}\left(\lambda_i^c\I{n}-A\right)^n$,
are then spanned by  complex generalized eigenvectors $v_{i,1}^r,v_{i,2}^r,\dots,v_{i,k_c^i}^c$. In order to generate real invariant subspace from these complex eigenspaces, one can instead use the fact that for any $v_i^c\in \mathcal{E}_i^c$, its complex conjugate satisfies $\overline{v_i^c}\in \overline{\mathcal{E}_{\lambda_i^c}^n}:=\text{ker}\left(\overline{\lambda_i^c}\I{n}-A\right)^n$. Thus for a complex eigenvalue $\lambda_i^c$ and its corresponding eigenvector $v_i^c$, the space spanned by  $\{\text{Re}[{v_i^c}], \text{Im}[{v_i^c}] \}$ is a two-dimensional invariant subspace of $A$. 

In terms of the real Jordan form, let $J^c_1,\dots,J_{k_c}^{r}$ be the Jordan blocks corresponding to the complex conjugate eigenvalue pairs $\{\lambda_i^c, \overline{\lambda_i^c}\}$  for $i=1,\dots,k_c$. Then for  $V_i^c=[\mathcal{V}^c_1,\dots,\mathcal{V}^c_{k_c}]\in\Ri{n\times 2k_c}$ where $\mathcal{V}^c_i:=\big[\text{Re}[{v_{i,1}^c}],\text{Im}[{v_{i,1}^c}]\big]\in\Ri{n\times 2}$, one has $AV_i^c=V_i^cJ_i^c$, such that for any  positive  integer $\mu\le k_c$,  one  can  construct  a  real  invariant subspace of even dimension $2\mu$, spanned  by $\mathcal{V}_1^c,\mathcal{V}_1^c,\dots,\mathcal{V}_\mu^c$,. Hence pairs of complex conjugate eigenvalues  may only generate invariant subspaces of even dimension, from which the two following well-known statements can be concluded.

\begin{lemma}\label{statement:invaraintSSexs}
A  matrix $A\in \Ri{n\times n}$ has real invariant subspaces of all even dimensions less than $n$, while it has real invariant subspaces of odd dimensions if, and only if, it has at least one real eigenvalue.
\end{lemma}

\begin{lemma}\label{statement:invaraintSSodd}
If   $n$ is odd, then $A\in \Ri{n\times n}$ has real invariant subspaces of all dimensions less than $n$.
\end{lemma}

%%%%%%%%%%%%%%%%%%%%%%%%
\subsection{On constructing Floquet--Lyapunov factorizations}\label{app:consFLfac}
There are several ways of computing  real ($cT$-period) Floquet--Lyapunov (FL) factorization for  LTP systems. These are mainly grouped into either  direct- or indirect methods. In the case of direct approaches (see e.g.  \citenum{zhou2008classification}) one uses  knowledge of the state transition matrix  to find $L(\cdot)$ and $F$ directly from \eqref{eq:FLfact}. The existence of a real matrix $F$  in Theorem~\ref{th:realFLtrans}, for instance, then follows from the fact that the Monodromy matrix, $\mono_A$, is real; indeed, using Lemma 3 in  \citenum{zhou2008classification}, we  have  
\begin{equation}
    \mono_A^2=\overline{\mono_A}{\mono_A}=e^{T B}e^{T \overline{B}}=e^{T (B+\overline{B})}=e^{2T F}
    \label{eq:realPLF}
\end{equation}
 such that $F=(B+\overline{B})/2$ for some possibly complex matrix $B$.

 In the indirect approach suggested in  \citenum{montagnier2004control}, on the other hand, one assumes that  $F\in\Ri{n\times n}$ satisfying \eqref{eq:FLfact} is known for some $cT$-periodic matrix $L(t)$, such that \eqref{eq:Ldiff} or \eqref{eq:Linvdiff}
 can  be solved for either $L(t)$ or  $L\inv(t)$, respectively, using that $L(0)=L\inv(0)=\I{n}$.
Indeed, the converse  is also true \cite{yakubovich1975}.
\begin{corollary}
    If there exists a matrix $F\in\Ri{n\times n}$ and a  $cT$-periodic, nonsingular matrix function $L:\Rip\to\Ri{n\times n}$,  $L(0)=\I{n}$, satisfying the matrix differential equation \eqref{eq:Ldiff}, then \eqref{eq:FLfact} is a real, $cT$-periodic FL factorization of \eqref{eq:STM}.
\end{corollary}

One therefore  has two natural options for finding an FL factorization: 
\begin{itemize}
    \item[1)]  Integrate \eqref{eq:STM} to find the Monodromy matrix and then  obtain $F$  from \eqref{eq:realPLF}, such that $L(t)$ can  be found either  from \eqref{eq:FLfact} directly or by integrating  \eqref{eq:Ldiff}; 
    \item[2)]  Or as suggested in  \citenum{montagnier2004control}:   Find both $F$ and $L(t)$ simultaneously by solving \eqref{eq:Ldiff} as a  boundary value problem using $L(0)=L(cT)=L(t)Y$ for some $k\in\{1,2\}$, and by taking $\dot{F}=0$.
    \end{itemize}
See also references \citenum{sinha1996liapunov} and \citenum{castelli2013rigorous} for alternative ways of computing real factorizations.

%%%%%%%%%%%%%%%%%%%%%%%%%%%%%%
%%%% Underactuated system %%%%
%%%%%%%%%%%%%%%%%%%%%%%%%%%%%%
\subsection{Trajectory Planning for Underactuated Mechanical Systems using  Virtual Constraints }\label{sec:UndAcMechSys}
We will now briefly demonstrate how the virtual constraints-approach of
 \citenum{shiriaev2005constructive} (see also  \citenum{shiriaev2010transverse,shiriaev2008can}) can be  used  to plan periodic trajectories of underactuated mechanical systems \cite{spong1998underactuated,liu2013survey} with one degree of underactuation.

The equations of motion of such systems may be written on the  form:  \cite{spong1998underactuated}
\begin{equation}
    M(\gc)\DDgc+C(\gc,\Dgc)\Dgc+G(\gc)=\hat{B}\hat{\ac}. \label{eq:EL}
\end{equation}
Here $\gc\in  \Ri{n_q}$ are the generalized coordinates,  $\Dgc=\frac{d}{dt}\gc$ the generalized velocities,  $\sysstate:=[\gc;\Dgc]\in \Ri{n}$ with $n=2n_q$ is the state vector, $\hat{\ac}\in \Ri{m}$ denotes the $m=n_q-1$ control inputs,  $M(\cdot)$ is the symmetric, positive definite inertia matrix, $C(\cdot)$ consists of centrifugal and Coriolis terms, $G(\cdot)$ is a vector of potential forces, while the constant matrix $\hat{B}\in \Ri{n_q\times m}$ has full rank.  

\paragraph{Constructing a periodic trajectory:}
 {To plan} a periodic trajectory for \eqref{eq:EL}, we introduce the vector function   $\vc(\mgvc)=[\vcs_1(\mgvc);\dots;\vcs_{n_q}(\mgvc)]$. It consist of $\mathcal{C}^3$-smooth scalar \emph{synchronization functions} $\phi_i(\cdot)$  which are built up of a finite number of basis functions. The idea behind introducing $\vc(\theta)$ is to specify a specific synchronization of the generalized coordinates in terms of the scalar variable $\mgvc$, hereafter referred to as the \emph{motion generator} (MG). For simplicity, the MG $\theta$  will be assumed to correspond to one of the generalized coordinates, that is $\theta\equiv q_i$ for some $i\in\{1,\dots,n_q\}$.\footnote{The geometric relations $\gcs_1=\vcs_1(\mgvc),\dots,\gcs_{n_q}=\vcs_{n_q}(\mgvc)$, $\theta=\theta(\gc)$, are commonly referred to as \emph{ virtual (holonomic) constraints}  by the fact that they are not inherent physical constraints imposed on the system, but rather must be enforced and kept invariant by some control action.} {It is interesting to note that the  invariance of such a relation then corresponds to a (forced) \emph{strict mode} of the mechanical system; see Def. 3 in Reference \citenum{albu2020review}. }
 
 For an underactuated system, it is important to note that the time evolution of the MG cannot be any for a specific choice of the synchronization function $\vc(\cdot)$. 
 Indeed, suppose that for some choice of $\vc(\cdot)$ and $\theta$ the system of \eqref{eq:EL}  has a non-trivial, bounded, $T$-periodic trajectory   $(\gc_\nom(t),\hat{\ac}_\nom(t))=(\gc_\nom(t+T),\hat{\ac}_\nom(t+T))$ whose orbit $\nomorb$, as  defined by \eqref{eq:NomOrb}, admits a reparameterization 
\begin{equation}\label{eq:VCparametr}
    \gc_\nom(t)=\vc(\mgvc_\nom(t)),  \quad \Dgc_\nom(t)=\vc'(\mgvc_\nom(t))\dot{\mgvc}_\nom(t), \quad \forall t \in [0,T),
\end{equation}
 with $\vc'(\mgvc)=\frac{d}{d\mgvc}\vc\left(\mgvc\right)$. This implies that $\theta_\star(t)$ in \eqref{eq:VCparametr} must be a solution of the so-called \emph{reduced dynamics}.  
\begin{proposition}[Reduced Dynamics \cite{shiriaev2005constructive}] \label{prop:redDyn}
Assume the invariance of the relations $q=\Phi(\mgvc)$ in \eqref{eq:VCparametr}, and consequently also $\dot{q}=\Phi'(\mgvc)\dot{\mgvc}$ and $\ddot{q}=\Phi'(\mgvc)\ddot{\mgvc}+\Phi''(\mgvc)\dot{\mgvc}^2$. Then $\mgvc(t)$ is the solution of a  second-order differential equation of the form
\begin{equation}
    \alpha(\mgvc)\ddot{\mgvc}+\beta(\mgvc)\dot{\mgvc}^2+\gamma(\mgvc)=0
    \label{eq:RedDyn}
\end{equation}
in which the smooth scalar functions $\alpha(\cdot)$, $\beta(\cdot)$ and $\gamma(\cdot)$ are defined as
\begin{align*}
    \alpha(\mgvc):=\hat{B}^\perp M\big(\vc(\mgvc)\big)\vc'(\mgvc), 
    \
    \beta(\mgvc):=\hat{B}^\perp M\big(\vc(\mgvc)\big)\vc''(\mgvc)+\hat{B}^\perp C\big(\vc(\mgvc),\vc'(\mgvc)\big)\vc'(\mgvc), 
    \end{align*}
and $ \gamma(\mgvc):=\hat{B}^\perp G\big(\vc(\mgvc)\big)$, with $\hat{B}^\perp\in \Ri{1\times n}$ a full-rank left annihilator of $\hat{B}$, that is $\hat{B}^\perp\hat{B}=\0{1\times m}$. 
\end{proposition}

It  follows that any periodic solution $\mgvc_\nom(t)=\mgvc_\nom(t+T)$, $T>0$,  of the {reduced dynamics} \eqref{eq:RedDyn}, given a choice of $\Phi(\mgvc)$ and initial conditions $(\mgvc_0,\dot{\mgvc}_0)$, defines a periodic solution of \eqref{eq:EL}. 

Another important property of the reduced dynamics equation \eqref{eq:RedDyn}  is the fact that, if $\alpha(\mgvc)\neq 0$, then it is integrable, with the integrating factor given by $\alpha(\mgvc)\psi(\mgvc,\mgvc_0)$ where
\begin{equation*}
     \psi(\mgvc_0,\mgvc):=\exp{\left(-2\int_{\mgvc_0}^\mgvc\frac{\delta(\upsilon)}{\alpha(\upsilon)}  d\upsilon\right)}, \quad  \delta(\mgvc):=\beta(\mgvc)-\frac{d}{d\mgvc}\alpha(\mgvc).
\end{equation*}
From this, the following statement, which is a slight reformulation of theorems 1 and 2 in  \citenum{shiriaev2005constructive}, can be easily deduced.
\begin{proposition} \label {prop:intFunc}
Let $\mgvc_\nom(t)=\mgvc_\nom(t+T)$ be a bounded, $T$-periodic solution of \eqref{eq:RedDyn}  corresponding to the initial conditions $(\mgvc_0,\dot{\mgvc}_0)$ on which $\alpha(\mgvc)\neq 0$. Then the function $I=I(\mgvc,\dot{\mgvc},\mgvc_0,\dot{\mgvc}_0)$, defined by
\begin{equation}\label{eq:intFunc}
    I:=\frac{1}{2}\alpha^2(\mgvc)\dot{\mgvc}^2-\psi(\mgvc_0,\mgvc)\left[\frac{1}{2}\alpha^2(\mgvc_0)\dot{\mgvc}_0^2-\int_{\mgvc_0}^\mgvc \psi(\upsilon,\mgvc_0)\alpha(\upsilon)\gamma(\upsilon)d\upsilon \right]
\end{equation}
vanishes on the nominal orbit $\eta_\nom$. Moreover, 
\begin{equation}
    \frac{d}{dt}{I}=\dot{\mgvc}\left(\alpha(\mgvc)U-2\frac{\delta(\mgvc)}{\alpha(\mgvc)}I  \right)
    \label{eq:intDiff}
\end{equation} 
is the time derivative of $I(\cdot)$ along a solution $(\mgvc(t),\dot{\mgvc}(t))$ of  $\alpha(\mgvc)\ddot{\mgvc}+\beta(\mgvc)\dot{\mgvc}^2+\gamma(\mgvc)=U$.
\end{proposition}
Given a choice of synchronization functions and initial conditions $(\mgvc_0,\dot{\mgvc}))$, one can therefore find $\dot{\theta}$ (if it exists) from \eqref{eq:intFunc} for any value of $\theta$ instead of having to integrate \eqref{eq:RedDyn}.
%%%%%%%%%%%%%%
\paragraph{Choosing a regular parameterization and a projection operator:}
Suppose a periodic trajectory of the form \eqref{eq:VCparametr} has been found. The next step is then to obtain from it a parameterization of the form \eqref{eq:xsConds}.  That is to say, we need to find a $T$-periodic, strictly monotonically increasing scalar variable $\mg:[0,T)\to\mgspace$ and a $\cont{1}$-function $\nvel:\mgspace\to\Rip$ such that we have $(\mgvc_\star(\mg_\star(t)),\dot{\mgvc}_\star(\mg_\star(t)))$ and $\dot{\mg}_\nom(t)=\nvel(\mg_\nom(t))>0$ for all $t\in[0,T)$.  

For periodic trajectories where $\dot{\mgvc}
_\star>0$ for all $t\in[0,T)$, an obvious candidate is then to take $s=\mgvc$ and find $\nvel(\mgvc)=\dot{\mgvc}_\star(\mgvc)$ from \eqref{eq:intFunc}. However, this is not possible for solutions of \eqref{eq:RedDyn} which orbits an equilibrium point $\mgvc_e$ of  type center \cite{shiriaev2006periodic}, as then $\dot{\mgvc}_\star$ will also become negative along $\nomorb$. In these cases, one may pick  $\mg$ depending on the choice of projection operator (see Def.~\ref{def:ProjOp}). For example, an operator of the form  $(\mgvc,\dot{\mgvc})\mapsto \mg$ is given by Eqs. 8-9 in  \citenum{mohammadi2018dynamic}, while the following family of implicitly defined local operators
\begin{equation}\label{eq:GenProjOp}
    \prj(\sysstate)=\argmin_{\mg\in \mgspace} (\sysstate-\sysstate_\mg(\mg))\transp\Lambda(\mg)(\sysstate-\sysstate_\mg(\mg))
\end{equation}
where  $\Lambda:\mgspace \to \Ri{n\times n}$ is some smooth, symmetric PD matrix function\footnote{If $\frac{d}{dt}\xs(\mg)=f(\xs(\mg))$ (c.f.  \eqref{eq:NomNonLinSys}), then $\Lambda(\cdot)$ in fact only needs to be PSD and satisfy $f(\xs(\mg))\Lambda(\mg)f(\xs(\mg))>0$ for all $\mg\in\mgspace$.}, may be used for any  parameterization of the form \eqref{eq:xsConds}, including the time parameterization \eqref{eq:VCparametr}. 
%%%%%%%%%%%%%%%%%%%
\paragraph{Transverse Coordinates and the Linearized Transverse Dynamics} \label{sec:TVD}
Suppose now a desired trajectory \eqref{eq:VCparametr} has been found, together with a regular parameterization $\xs;\mgspace\to\nomorb $ and a projection operator $\prj(\cdot)$  (e.g. as \eqref{eq:CPparamAndProjOp} for the cart-pendulum). The next step is then to find a set of transverse coordinate (see Def.~\ref{def:TVC}) and to linearize their dynamics.
To this end, we will assume in the following that the MG may be taken as $\mgvc:=\gcs_{n_q}$. This assumption allows us to define the $(n-1)$ transverse coordinates $ \tvc=[y;\dot{y};I]$,  where  $y\in\Ri{n_q-1}$ and its derivative $\dot{y}\in\Ri{n_q-1}$ are defined as
 \begin{equation}\label{eq:ytvcDef}
     y:=L_y\left(\gc-\vc(\mgvc)\right), \quad \dot{y}:=L_y\left(\Dgc-\vc'(\mgvc)\dot{\mgvc}\right), \quad  L_y:=[
     \I{n_q-1}, \0{m\times 1}],
 \end{equation}
and with $I$ given by \eqref{eq:intFunc}. Indeed, $\tvc$ evidently vanishes along the solution \eqref{eq:VCparametr}, while the corresponding Jacobian matrix 
\begin{equation*}
    \jac{\tvc}(\sysstate)=\begin{bmatrix} \I{n_q-1} & -L_y\vc'(\mgvc) &  \0{n_q-1} & \0{(n_q-1)\times 1}\\
    \0{n_q-1} & -L_y\vc''(\mgvc)\dot{\mgvc} & \I{n_q-1} & -L_y\vc'(\mgvc)
    \\
    \0{1\times (n_q-1)} & {\partial_{\mgvc} I} & \0{1\times (n_q-1)} & \alpha^2(\mgvc)\dot{\mgvc}
    \end{bmatrix}, \ {\partial_{\mgvc} I}=\alpha(\mgvc)(\beta(\mgvc)\dot{\mgvc}^2+\gamma(\mgvc))-\frac{2\delta(\mgvc)}{\alpha(\mgvc)}I,
\end{equation*}
has (full) rank equal to $2n_q-1=n-1$ when evaluated along $\nomorb$ (simply note that $\partial_{\mgvc} I(\sysstate_\nom)=-\alpha^2(\mgvc_\nom)\ddot{\mgvc}_\nom$.  

Before we can linearize the dynamics of $\tvc$, we first need to rewrite \eqref{eq:EL} on a form similar to \eqref{eq:NonLinSys} for which $\dot{\sysstate}_\nom=f(\sysstate_\nom)$. We therefore introduce the smooth mappings $w:\Ri{n}\to\Ri{m}$  and $W:\Ri{n}\to\Ri{m\times m}$ which are such that, for all $t\in[0,T)$, one has $w(\sysstate_\nom(t))\equiv \hat{\ac}_\nom(t)$ and  $W(\sysstate_\nom(t))$ is nonsingular. Taking, therefore, 
\begin{equation}\label{eq:FeedbackTrans}
    \hat{\ac}=w(\sysstate)+W(\sysstate)\ac
\end{equation}
 one can rewrite \eqref{eq:EL} on the (disturbance-free) form of \eqref{eq:NomNonLinSys} 
with
\begin{equation*}
    f(\sysstate)=\begin{bmatrix} \dot{q} \\ M\inv(\gc)\left[\hat{B}w(\sysstate)-C(\gc,\Dgc)\Dgc-G(\gc)\right] \end{bmatrix}, \quad g(\sysstate)=\begin{bmatrix} \0{n\times m} \\ M\inv(\gc)\hat{B}W(\sysstate) \end{bmatrix}.
\end{equation*} 
Consequently, the linearized transverse dynamics may then be found  using Lemma~\ref{lemma:TVL}, such that a nominal state feedback controller can be designed utilizing, for example, Proposition~\ref{prop:PRDE}.

%%%%%%%%%%%%%%%%%%%%%%%%%%%%
%%%% Proofs %%%%%%%%%%%%%%%%
%%%%%%%%%%%%%%%%%%%%%%%%%%%%
\section{Proofs of statements}\label{app:proofs}
%%%%%%%%%%%%%%%%%%%%%
% \subsection{Proof of Lemma~\ref{lemma:linCLsys}}\label{app:Lemma2proof}

%%%%%%%%%%%%%%%%%%%
% \subsection{Proof of Proposition~\ref{prop:LinCL}}

%%%%%%%%%%%%%%%%%%%%%%%%%%%%%%%%%%%
\subsection{Proof of Lemma~\ref{lemma:LTPstatement}}\label{proof:lemma:LTPstatement}
% \begin{proof}
    First note that  at each time $t\in\Rip$ one must have 
    \begin{equation*}
        \frac{d}{dt}\left[S(t)\stm_{A^{cl}}(t,0)X_0\right]=\dot{S}(t)\stm_{A^{cl}}(t,0)X_0+S(t){A^{cl}}(t)\stm_{A^{cl}}(t,0)X_0=0 
    \end{equation*}
    which implies the relation $\dot{S}(t)\stm_{A^{cl}}(t,0)X_0=-S(t){A^{cl}}(t)\stm_{A^{cl}}(t,0)X_0$.
     
     Now assuming the forward invariance of $S(t)y(t)=0$,  it follows that
    \begin{equation*}
        \frac{d}{dt}\left(S(t)y(t)\right)=\dot{S}(t)\linstate+S(t)\left(A(t)\linstate+B(t)(\ac_{eq}+\dist)\right)\equiv 0.
    \end{equation*}
    Therefore, as if $S(t)\linstate(t)=0$ then $\linstate(t)=\stm_{A^{cl}}(t,0)X_0p$ for some $p\in\Ri{\bar{n}-\bar{m}}$, we obtain, using the above relation,
    \begin{equation*}
        -S(t){A^{cl}}(t)\linstate+S(t)\left(A(t)\linstate+B(t)(\ac_{eq}+\dist)\right)=S(t)B(t)\left[\ac_{eq}+\dist-\lfb(t)\linstate\right]\equiv0.
    \end{equation*}
     Due to the assumption that $\rank[S(t)B(t)]=\bar{m}$, the  equivalent control  is then uniquely given by $\ac_{eq}=\lfb(t)\linstate-\dist$.
    
    To derive the stated condition for the $T$-periodicity of $S(\cdot)$,  recall the following property of the STM  \cite{hale1971functional}: $\stm_{A^{cl}}(t+T,0)=\stm_{A^{cl}}(t,0)\mono_{A^{cl}}$  . Thus if $S(t)=S(t+T)$ for all $t\ge 0$, then
    \begin{equation*}
        \|S(t){\stm}_{A^{cl}}(t,0)X_0p\|=\|S(t+T){\stm}_{A^{cl}}(t+T,0)X_0p\|=\|S(t){\stm}_{A^{cl}}(t,0)\mono_{A^{cl}}X_0p\| \equiv 0
    \end{equation*}
     for any $p\in\Ri{(\bar{n}-\bar{m})}$ and all $t\ge 0$. It follows that there must be some nonsingular matrix $N\in\Ri{(\bar{n}-\bar{m})\times (\bar{n}-\bar{m})}$ such that $\mono_{A^{cl}}X_0=X_0N$, or equivalently, the columns of $X_0$ form a basis of an invariant subspace of $\mono_{A^{cl}}$. 
% \end{proof}
%%%%%%%%%%%%%%%%%%%%%%%%
\subsection{Proof of Proposition~\ref{prop:SMdesignLTP}}\label{proof:prop:SMdesignLTP}
% \begin{proof}
    Assume,  without loss of generality, that the  relation 
     $\sigma(t)=S(t)\linstate(t)\equiv 0 $
    {is forced for $t\ge 0$.  Following the equivalent control approach,\cite{utkin2013sliding,shtessel2014sliding} we then assume that, for all $t\ge 0$,}
    \begin{equation*}
    \dot{\sigma}(t)=\hat{S}\left[\frac{d L\inv}{dt}(t)\linstate+L\inv(t)\left(A(t)\linstate+B(t)(\ac_{eq}+\dist)\right)\right]\equiv 0.
    \end{equation*}
    Here the derivative $\frac{d}{dt}L\inv(t)$ can be determined by first substituting \eqref{eq:FLfact} into \eqref{eq:STM}  in order to  obtain
\begin{equation}
    \frac{d}{dt}L(t)=A^{cl}(t)L(t)-L(t)F.
    \label{eq:Ldiff}
\end{equation}
Since $\frac{d}{dt}L\inv(t)=-L\inv(t)\dot{L}(t)L\inv(t)$ for any smooth,  nonsingular square matrix function  $L(\cdot)$, it therefore follows that
\begin{equation}
    \frac{d}{dt}L\inv(t)=-L\inv(t)A^{cl}(t)+FL\inv(t).
    \label{eq:Linvdiff}
\end{equation}
    Hence the above reduces to
    \begin{align*}        \dot\smf&=\hat{S}\left[FL\inv(t)\linstate-L\inv(t)\big(A^{cl}(t)-A(t)\big)\linstate+L\inv(t)B(t)(\ac_{eq}+\dist)\right] \nonumber
        \\
        &=\hat{S}\left[Fz-L\inv(t)B(t)\lfb(t)\linstate+L\inv(t)B(t)(\ac_{eq}+\dist)\right] \nonumber
        \\
        &=\hat{S}Fz+{S}(t)B(t)\left[\ac_{eq}-\lfb(t)\linstate+\dist\right]\equiv 0. 
    \end{align*}
    Now, as $\Lambda$ is $F$-invariant and $\hat{S}$ annihilates $\Lambda$,  we here have that $\hat{S}z=\hat{S}Fz\equiv 0$ for all $z\in\Lambda$.    Therefore, as we have assumed that $\left[S(t)B(t)\right]$ is invertible, it  follows that the equivalent control corresponds to $\ac_{eq}(t)=\lfb(t)\linstate(t)-\dist(\linstate(t),t)$ as desired.
% \end{proof}
%%%%%%%%%%%%%%%%%%%
\subsection{Proof of Lemma~\ref{lemma:IntroStatement}}\label{proof:lemma:IntroStatement}
% \begin{proof}
    In order to show convergence to $\nomorb$ within some nonzero tubular neighbourhood when restricted to $\SM$, 
     let $\tau:\Ri{n}\to [0,T)$ denote  the solution to the implicit equation
    \begin{equation}\label{eq:tauProj}
        \tau(\sysstate)=\argmin_{\tau\in[0,T)}\|\sysstate-\sysstate_\nom(\tau)\|^2,
    \end{equation}
     for a given $\sysstate\in\Ri{n}$ about $\nomorb$.  By similar arguments as for example those given in the work of Leonov \cite{leonov2006generalization}, the time derivative of $\tau=\tau(\sysstate)$ is well defined in a neighbourhood of $\nomorb$ and its dynamics may be written as \cite{hauser1994converse,saetre2020excessive}
    \begin{equation*}
        \dot{\tau}=1+f_\parallel(\tau,\tilde{\sysstate})  +g_\parallel(\tau,\tilde{\sysstate}) \ac
    \end{equation*}
    where   $\tilde{\sysstate}:=\sysstate-\sysstate_\nom(\tau)$  and  $f_\parallel(\tau,0)=0.$
    Any $\cont{2}$-function $\sysstate\mapsto h(\sysstate)$ may then be written on the form
    \begin{equation*}
        h(\sysstate)=h(\sysstate_\nom(\tau))+\jac{h}(\sysstate_\nom(\tau))\tilde{\sysstate}+R_h({\sysstate}) 
    \end{equation*}
    for $\sysstate$ in a small neighbourhood of $\nomorb$,
    where $R_h$ is  $\mathcal{C}^1$ and satisfies  $\|R_h({\sysstate})\|=\bigO(\|\tilde{\sysstate}\|^2)$.
    We can therefore  take
    \begin{align*}
        \smf(\sysstate)=S(\tau)\tilde{\sysstate}+R_{\smf}({\sysstate}) ,
        \quad 
        f(\sysstate)=f(\sysstate_\nom(\tau))+A(\tau)\tilde{\sysstate}+R_{f}({\sysstate})
        \quad \text{and} \quad
        \nfb(\sysstate)=\lfb(\tau)\tilde{\sysstate}+R_{k}({\sysstate}),
    \end{align*}
    where  we have used that $\smf(\sysstate_\nom(t))=\nfb(\sysstate_\nom(t))\equiv 0$. Hence, by differentiating $\smf(\cdot)$ with respect to time, one obtains
    \begin{align*}
        \dot{\smf}(\sysstate)&=\dot{S}(\tau)\tilde{\sysstate}+S(\tau)(\dot{\sysstate}-\dot{\sysstate}_\nom(\tau))+\frac{d}{dt}R_{\smf}({\sysstate}) 
        \\
        &=\dot{S}(\tau)\tilde{\sysstate}+S(\tau)\big(f(\sysstate)+g(\sysstate)\big[\ac+\dist\big]\big)+\frac{d}{dt}R_{\sigma}({\sysstate})
        \\
        &=\left[\dot{S}(\tau)+S(\tau)A(\tau)\right]\tilde{\sysstate}+S(\tau)g(\sysstate)\big[\ac+\dist\big]+\hat{R}
        \\
        &=\left[\dot{S}(\tau)+S(\tau)A(\tau)\right]\tilde{\sysstate}+S(\tau)\left(B(\tau)+\tilde{g}(\sysstate)\right)\big[\ac+\dist\big]+\hat{R}
        ,
    \end{align*}
    where $\tilde{g}(\sysstate):=g(\sysstate)-g(\sysstate_\nom(\tau))$, $\hat{R}:=S(\tau)R_f+\frac{d}{dt}R_{\smf}$, and where we have used that $S(\tau)\dot{\sysstate}_\nom(\tau)=S(\tau)\dot{\tau}\frac{d}{d\tau}\sysstate_\nom(\tau)\equiv 0$. 
    {By adding and subtracting $S(\tau)B(\tau)\lfb(\tau)\tilde{\sysstate}$, the above may be equivalently  rewritten as}
    \begin{equation*}
       { \dot{\smf}(\sysstate)=\left[\dot{S}(\tau)+S(\tau)A^{cl}(\tau)\right]\tilde{\sysstate}-S(\tau)(B\tau)\lfb(\tau)\tilde{\sysstate}+S(\tau)\big(B(\tau)+\tilde{g}(\sysstate)\big)\big[\ac+\dist\big]+\hat{R}. }
    \end{equation*}
    
    With this in mind, suppose $\SM$ is rendered forward invariant such that $\smf(\sysstate)=S(\tau)\tilde{\sysstate}+R_{\sigma}({\sysstate})\equiv 0$ and consequently also $\dot{\smf}(\sysstate)\equiv 0$ {following the equivalent control approach\cite{utkin2013sliding}}. {This implies that} $\tilde{\sysstate}=X_S(\tau)-S\pinv(\tau)R_{\sigma}({\sysstate})$ for some $X_S(\tau)\in\ker\{S(\tau)\}$ and in which $S\pinv(\tau)$ is a right inverse of $S(\tau)$  such that $S(\tau)S\pinv(\tau)=\I{m}$ for all $\tau\in[0,T)$. 
    Using condition 2. in Lemma~\ref{lemma:IntroStatement}, the equivalent control, $\ac_{eq}$, must therefore satisfy
    \begin{align*}
        \left[\I{m}+\left(S(\tau)B(\tau) \right)\inv S(\tau)\tilde{g}(\sysstate)\right]&(\ac_{eq}+\dist)
        \\
        &=\lfb(\tau)\tilde{\sysstate}- \left(S(\tau)B(\tau) \right)\inv\left[\left(\dot{S}(\tau)+S(\tau)A^{cl}(\tau)\right)S\pinv(\tau)R_{\sigma}({\sysstate})+\hat{R}\right]
    \end{align*}
    Since we have assumed the columns of  $g(\cdot)$ to be be locally Lipschitz in some region containing the orbit, there necessarily exists     a Lipschitz constant $\lipschitz_g>0$ such that $\|\tilde{g}(\sysstate)\|\le \lipschitz_g \|\tilde{\sysstate}\|$ holds therein. 
    {It follows that for sufficiently small $\tilde{\sysstate}$, the matrix function $\Lambda(\sysstate):=\I{m}+\left(S(\tau)B(\tau) \right)\inv S(\tau)\tilde{g}(\sysstate)$ is nonsingular. This, in turn, implies that, locally, it is of the form $\ac_{eq}=\hat{\nfb}(\sysstate)-\dist$ with
    \begin{equation*}
        \hat{\nfb}(\sysstate):=\Lambda\inv(\sysstate)\left(\lfb(\tau)\tilde{\sysstate}- \left(S(\tau)B(\tau) \right)\inv\left[\left(\dot{S}(\tau)+S(\tau)A^{cl}(\tau)\right)S\pinv(\tau)R_{\sigma}({\sysstate})+\hat{R}\right]\right). 
    \end{equation*}
    What remains is therefore to show that $\hat{\nfb}(\cdot)$ is  equal to $\nfb(\cdot)$ in the first approximation along $\nomorb$. Indeed, if this is the case, then necessarily $\ac_{eq}=-\dist(\sysstate_\nom(t),t)$ on $\nomorb\subset \SM$,  illustrating the insensitivity to the matched disturbance.}
    
    {To this end, we first  note that $\Lambda\inv(\sysstate_{\star}(t))= \I{m}$ for all $t\in[0,T)$. Moreover, since the Jacobian matrix of $\tilde{\sysstate}$ evaluated along the nominal motion is given by\cite{leonov2006generalization,saetre2020excessive}  $D\tilde{\sysstate}(\sysstate_\nom(t))=\I{n}-{\dot{\sysstate}_\nom(t)\dot{\sysstate}_\nom\transp(t)}/{\|\dot{\sysstate}_\nom(t)\|^2}$, as well as that $\|\lfb(t)\dot{\sysstate}_\star(t)\|\equiv 0$ for all $t\in[0,T)$, the relation $\lfb(t)D\tilde{\sysstate}(\sysstate_\nom(t))\equiv \lfb(t)$ always  holds. 
    Since all the terms inside the brackets on the right-hand inside of the expression for $\hat{\nfb}(\cdot)$ are necessarily of order no less than two with respect to $\tilde{\sysstate}$ as $\|\tilde{\sysstate}\|\to 0$, we can conclude that $D\hat{k}(\sysstate_\nom(t))\equiv \lfb(t)$, and thus $\hat{k}(\cdot)$ equals $\nfb(\cdot)$ in the first-order approximation as desired.}

% \end{proof}
%%%%%%%%%%%%%%%%
\subsection{Proof of Theorem~\ref{theorem:mainResult}}\label{proof:theorem:mainResult}
% \begin{proof}
    It is here enough to show that \eqref{eq:sigmaInMainTheorem} satisfies the requirements in Lemma~\ref{lemma:IntroStatement}. In this regard,  $\|\sigma(\xs(\mg))\|\equiv 0$ is trivially satisfied as $\|\tvc(\xs(\mg))\|\equiv 0$,  whereas $\jac{\sigma}(\xs(\mg))g(\xs(\mg))=S_\perp(\mg)\jac{\tvc}(\xs(\mg))g(\xs(\mg))=S_\perp(\mg)B_\perp(\mg)$ demonstrates that $\rank[S(t)B(t)]=m$ holds as well.
    
    In order to show that condition 2. in Lemma~\ref{lemma:IntroStatement} also holds for any $\sysstate\in\ker\{S(\mg(t))\}$, we  note that  \eqref{eq:Linvdiff} for  \eqref{eq:TVLCL} may be written as
    \begin{equation}\label{eq:LinvTVC}
        \nvel(\mg)\frac{d}{d\mg}L\inv(\mg)=-L\inv(\mg)A_\perp^{cl}(\mg)+FL\inv(\mg),
    \end{equation}
    where we recall that $ \nvel(\mg) :=\|f(\xs(\mg))\|/\|\nomflow(\mg)\|$ is such that $\dot{\mg}=\nvel(\mg)$ and $\nomflow(\mg):=\frac{d}{d\mg}\xs(\mg)$.
    We therefore obtain
    \begin{align*}     \dot{S}(\mg(t))=\frac{d}{dt}\left(\hat{S}L\inv(\mg)\jac{\tvc}(\xs(\mg))\right)
=&S_\perp(\mg)\Big[\big(-A_\perp^{cl}(\mg)+L(\mg)FL\inv(\mg)\big)\jac{\tvc}(\xs(\mg))
\\
&+\nvel(\mg)\frac{d}{d\mg}\jac{\tvc}(\xs(\mg))\Big].
    \end{align*}

    Now, since $S(\mg)=\hat{S}L\inv(\mg)\jac{\tvc}(\xs(\mg))$, it follows that  $\sysstate\in\ker\{S(\mg)\}$ corresponds to either $\sysstate\in\text{span}\{f(\xs(\mg))\}$ or $\sysstate=(\jac{\tvc})\pinv(\xs(\mg))L(\mg)(\I{n}-\hat{S}\pinv\hat{S})\sysstate$. Taking therefore $\sysstate=f(\xs(\mg))$ and using that $\jac{\tvc}(\xs(\mg))f(\xs(\mg))\equiv 0$, it is easy to see that condition 2. in Lemma~\ref{lemma:IntroStatement} is satisfied as
    \begin{equation*}
        \frac{d}{dt}\left[\jac{\tvc}(\xs(\mg))f(\xs(\mg))\right]=\nvel(\mg)\left[\frac{d}{d\mg}\jac{\tvc}(\xs(\mg))\right]f(\xs(\mg))+\jac{\tvc}(\xs(\mg))A^{cl}(\mg)f(\xs(\mg))=0.
    \end{equation*}
    Hence we need only demonstrate that the following always holds:
    \begin{equation}\label{eq:satisfySubspaceProp}
        \left[\dot{S}(\mg)+S_\perp(\mg)\jac{\tvc}(\xs(\mg))A^{cl}(\mg) \right](\jac{\tvc})\pinv(\xs(\mg))L(\mg)(\I{n}-\hat{S}\pinv\hat{S})=\0{n}.
    \end{equation}

    From the definition of the matrix function $A_\perp^{cl}$ (see \eqref{eq:TVLCL}), it can be shown that 
    \begin{equation*}
        A_\perp^{cl}(\mg)=\left[\jac{\tvc}(\xs(\mg))A^{cl}(\mg)+\nvel\frac{d}{d\mg}\jac{\tvc}(\xs(\mg)) \right](\jac{\tvc})\pinv(\xs(\mg)).
    \end{equation*}
    Using this together with the above expression for $\dot{S}(\mg)$, \eqref{eq:satisfySubspaceProp} therefore reduces to
    \begin{equation*}
        S_\perp(\mg)\left[L(\mg)FL\inv(\mg)\jac{\tvc}(\xs(\mg))\right](\jac{\tvc})\pinv(\xs(\mg))L(\mg)(\I{n}-\hat{S}\pinv\hat{S})=\hat{S}F(\I{n}-\hat{S}\pinv\hat{S})=0 
    \end{equation*}
    where we have used that $\hat{S}Fz\equiv 0$ for all $z\in\ker\{\hat{S}\}=\{z\in\Ri{n}: \ z=(\I{n}-\hat{S}\pinv\hat{S})z\}$, This concludes the proof.\footnote{Alternatively, the statement can be proven by utilizing the fact that  \eqref{eq:TVLCL} is real reducible (the existence of a real FL factorization has been assumed) in order to invoke  Theorem 25 in the work of Massera \cite{massera1956contributions}. This allows one to conclude that the origin of the transverse dynamics  \eqref{eq:TVD} is locally asymptotically stable when in sliding mode,  and, therefore, by Proposition 1.5 from Hauser and Chung \cite{hauser1994converse}, that the   solution $\sysstate_\nom(\cdot)$  is exponentially orbitally stable. 
    }
% \end{proof}
%%%%%%%%%%%%%
\subsection{Proof of Lemma~\ref{lemma:PreContDesign}}\label{proof:lemma:PreContDesign}
% \begin{proof}
    Firstly, since the nominal exponentially orbitally stabilizing feedback $k(\cdot)$  is $\cont{2}$ and satisfies $k(\sysstate_\nom)\equiv 0$, one may write $k(\sysstate)=K_\perp(\mg) \tvc+R_k(\tvc,\mg)$ in which  $\|R_k(\tvc,\cdot)\|=\bigO(\|\tvc\|^2)$. Using this, together with  Lemma~\ref{lemma:TVL} and the fact that $f_\perp(\cdot)$ is continuously differentiable (as $f(\cdot)$ and   $\tvc(\cdot)$ are assumed to be  $\cont{2}$)  we may  then rewrite \eqref{eq:TVD} on  form: \begin{equation}\label{eq:TVD2}
        \frac{d}{dt}{\tvc}=A_\perp^{cl}(\mg)\tvc+\big[B_\perp(\mg)+\tilde{g}_\perp(\tvc,\mg)\big]\left(v+\dist(\sysstate,t)\right)+R_\perp, \quad \mg=\prj(\sysstate).
    \end{equation}
     Here  $R_\perp:=R_{f_\perp}(\tvc,\mg)+g(\sysstate)R_k(\tvc,\mg)$  and $\tilde{g}_\perp:=g_\perp(\tvc,\mg)-B_\perp(\mg)$ satisfy, respectively, 
        $\|R_{\perp}(\tvc,\mg)\|=\bigO(\|\tvc\|^2)$ and   $\|\tilde{g}_\perp(\tvc,\mg)\|\le \lipschitz_{\tilde{g}_\perp}\|\tvc\|$
    for all $\mg\in\mgspace$, with $\lipschitz_{\tilde{g}_\perp}> 0$ a Lipschitz constant for $\tilde{g}_\perp$, whose (local) existence is guaranteed as $\jac{\tvc}$ is $\cont{1}$ and the columns of $g(\cdot)$ are locally Lipschitz. 
    Differentiating \eqref{eq:sigmaInMainTheorem} with respect to time and using \eqref{eq:LinvTVC}, we therefore  obtain
         \begin{align*}
           \dot{\sigma}&=\dot{S}_\perp(\mg)\tvc+S_\perp(\mg)\left[A_\perp^{cl}(\mg)\tvc+\big[B_\perp(\mg)+\tilde{g}_\perp(\tvc,\mg)\big]\left(v+\dist\right)+R_\perp\right] 
           \\ \nonumber
           &=\hat{S}\left[-L(\mg)\inv A_\perp^{cl}(\mg)    +FL\inv(\mg)\right]\frac{\dot{s}}{\nvel(\mg)}\tvc +S_\perp(\mg)\left[A_\perp^{cl}(\mg)\tvc+\big[B_\perp(\mg)+\tilde{g}_\perp(\tvc,\mg)\big]\left(v+\dist\right)+R_\perp\right].
           \end{align*}
    As $\dot{\mg}=\Dp(\sysstate)\dot{\sysstate}$ and $\nvel(\mg)=\Dp(\xs(\mg)f(\xs(\mg))=f(\xs(\mg))/\|\nomflow(\mg)\|$, we may here, in the same manner as with \eqref{eq:TVD2}, take
    \begin{equation}\label{eq:DmgOmega}
        \dot{\mg}=\nvel(\mg)+\jac{f}_\parallel(\xs(s)){\big(\I{n}-\nomflow(\mg)\jac{\prj}(\xs(\mg))\big)}(\jac{\tvc})\pinv(\xs(\mg))\tvc+g_{\parallel}(\sysstate)(v+\dist(\sysstate,t))+R_{f_\parallel}(\tvc,\mg),
    \end{equation}
    where $\mg=\prj(\sysstate)$, $f_\parallel(\sysstate):=\Dp(\sysstate)f(\sysstate)$ and $g_\parallel(\sysstate):=\Dp(\sysstate)g(\sysstate)$, while $\|R_{f_\parallel}(\tvc,\mg)\|=\bigO(\|\tvc\|^2)$; see Reference \citenum{mohammadi2018dynamic} or \citenum{saetre2020excessive} for more details.
    
    Since one can always write $\tvc=L(\mg)\hat{S}\pinv\sigma+(\I{n-1}-L(\mg)\hat{S}\pinv S_\perp(\mg))\tvc$, as well as as that $\hat{S}F(\I{n-1}-\hat{S}\pinv\hat{S})\equiv 0$, it follows from the expression above that $\dot{\sigma}$ can be written on the form \eqref{eq:sigmaDot} with 
    \begin{align*}
        \tilde{B}_\perp(\tvc,\mg)&=\tilde{g}_\perp(\tvc,\mg)+L(\mg)\left(\frac{d}{ds}L\inv(\mg)\right)\tvc g_\parallel(\sysstate),
        \\
        R_\sigma(\tvc,\mg)&=S_\perp(\mg)\Big[R_\perp(\tvc,\mg)+L(\mg)\left(\frac{d}{ds}L\inv(\mg)\right)\tvc R_{f_\parallel}(\tvc,\mg)
        \\
        &+L(\mg)\left(\frac{d}{ds}L\inv(\mg)\right)\tvc\left( \jac{f}_\parallel(\xs(\mg)){\big(\I{n}-\nomflow(\mg)\jac{\prj}(\xs(\mg))\big)}(\jac{\tvc}))\pinv(\xs(\mg))\tvc \right)\Big].
    \end{align*}
    
    Lastly, since $\hat{S}$ annihilates a stable invariant subspace of $F$, spanned by a set of its (real) generalized eigenvectors, $\mathcal{F}_\sigma$ must necessarily be Hurwitz with its spectrum  a subset of the spectrum of $F$; see the proof of Lemma~\ref{lemma:linCLsys} for more details.
% \end{proof}

\end{document}